\documentclass[pre,notitlepage,showpacs,showkeys,longbibliography]{revtex4-1}
\usepackage{latexsym,amssymb,amsfonts,mathrsfs,amsmath,bm}
\usepackage{graphicx}
\usepackage[usenames,dvipsnames]{xcolor}
\usepackage{soul}
\usepackage{float}
\usepackage[linktocpage,colorlinks=true,linkcolor=blue,citecolor=blue,breaklinks=true,urlcolor=blue]{hyperref}
\usepackage{textcomp}

\newcommand{\gammadot}{{\dot\gamma}}
\newcommand{\bA}{{\bf A}}

\newcommand{\bE}{{\bf E}}
\newcommand{\bI}{{\bf I}}
\newcommand{\bG}{{\bf G}}
\newcommand{\bK}{{\bf K}}

\newcommand{\bR}{{\bf R}}
\newcommand{\bX}{{\bf X}}
\newcommand{\bU}{{\bf U}}
\newcommand{\bV}{{\bf V}}

\newcommand{\be}{{\bf e}}
\newcommand{\bff}{{\bf f}}

\newcommand{\bu}{{\bf u}}
\newcommand{\bx}{{\bf x}}
\newcommand{\by}{{\bf y}}

\newcommand{\bzero}{{\bf 0}}
\newcommand{\barbu}{\bar{\bu}}

\usepackage{amsmath}
\usepackage{xcolor}
\usepackage{units}
\usepackage{natbib}

\begin{document}


\title{Dynamics of flexible fibers in viscous flows and fluids}

\author{O. du Roure,$^1$}
\author{A. Lindner,$^{1,2}$}
\author{E.N. Nazockdast$^3$}
\author{M.J. Shelley$^{4,5}$}
\date{\today}

\affiliation{$^1$ Laboratoire de Physique et M´ecanique des Milieux H´et´erog`enes (PMMH),
ESPCI Paris, PSL University, CNRS, Sorbonne University, and Paris Diderot
University, 75005 Paris, France; email: olivia.duroure@espci.fr}
\affiliation{$^2$ Global Station for Soft Matter, Global Institution for Collaborative Research
and Education, Hokkaido University, Sapporo 060-0808, Japan}
\affiliation{$^3$ Applied Physical Sciences, University of North Carolina at Chapel Hill, North
Carolina 27514, USA}
\affiliation{$ 4$ The Courant Institute of Mathematical Sciences, New York University, New
York, NY 10012, USA}
\affiliation{$ 5$ Center for Computational Biology, Flatiron Institute, New York, NY 10010,
USA}

\begin{abstract}
The dynamics and deformations of immersed flexible fibers are at the
heart of important industrial and biological processes, induce
peculiar mechanical and transport properties in the fluids that
contain them, and are the basis for novel methods of flow
control. Here we focus on the low Reynolds number regime where
advances in studying these fiber-fluid systems have been especially
rapid. On the experimental side this is due to new methods of fiber
synthesis, microfluidic flow control, and of microscope based
tracking measurement techniques. Likewise, there have been continuous
improvements in the specialized mathematical modeling and numerical
methods needed to capture the interactions of slender flexible fibers
with flows, boundaries, and each other.
\end{abstract}

\maketitle

%

\section{INTRODUCTION AND BACKGROUND}
\label{sec:Introduction}

An important class of micro-scale fluid-structure interactions involves
the interactions and deformations of flexible fibers with fluid
flows. This is evident in the many biological transport processes,
such as microorganismal swimming (\cite{Lauga2009}) or nuclear
positioning in eukaryotic cells (\cite{Shelley2016}), that involve
flexible fibers either actuated or passive. Fibers are the
microstructure of many complex fluids -- biological, industrial, and
synthetic -- studied both for their scientific and industrial
importance and for their peculiar mechanical responses
(\cite{Broedersz2014}). Such suspensions are particularly challenging
to study because the suspended fibers have many degrees of freedom in
deformation, and can exhibit microscopic instabilities. This makes the
interaction of fibers with background flows surprisingly
complex. Flexible fibers, both anchored and freely suspended, have
been studied of late for the purposes in microfluidic flow control,
and when actuated can exhibit complex collective and transport
dynamics.

This review focuses on the low Reynolds number
regime, as progress on fiber-fluid interactions in this regime has
been especially rapid. On the experimental side, this is due to
improvements in fiber synthesis and characterization, microfluidic flow control and improved
microscope-based measurement and particle tracking techniques. By
taking advantage of the relative simplicity of the Stokes equations
(as opposed to Navier-Stokes), combined with adaptive resolution and fast
summation approaches, there has been a likewise rapid improvement
in numerical methods for simulating the deformations and interactions
of fibers with flows, as well as with each other and other immersed
structures. Coarse-grained descriptions of fiber assemblies and
suspensions remain in their early stages, but their development is
being sped by these new numerical methodologies.

In outline, we first discuss the current state-of-art in experimental
synthesis and measurement techniques, and in numerical methods
for dynamical simulation fiber-fluid interactions. We then describe
the results of both experiment and theory for the dynamics of free
and anchored fibers, followed by a review of the current state of
research in many-fiber/fluid systems. We close with a discussion
of future directions.

\subsubsection{Setting the stage}

Here we briefly introduce a few important physical parameters and
relations. Consider a slender elastic fiber of length $L$, of circular
cross-section with radius $a$ (hence $\epsilon=a/L \ll 1$), and
flexural rigidity $E=YI$ with $Y$ the material Youngs modulus and $I$
the areal moment of inertia ($I=\pi a^{4}/4$). This fiber is immersed
in a Newtonian fluid of shear viscosity $\mu$ with the fluid motion
characterized by a strain-rate $\gammadot$. Neglecting inertial
forces, three important forces are at play: Brownian forces $\sim k_B
T/L$ ($k_B$ is the Boltzmann constant and $T$ the temperature), drag
forces $\sim\mu\gammadot L^2$, and elasticity forces $\sim E/L^2$. For
most of the work reviewed here, though not all, viscous drag and
elasticity forces dominate Brownian forces. That predominance requires
that $l_p/L \gg 1$ and $Pe =8\pi \mu \dot{\gamma}L^3/k_b T \gg 1$,
where $l_p=E/k_bT$ is the persistence length of the fiber against
thermal fluctuations, and the P\'eclet number, $Pe$, is the ratio of
viscous to Brownian forces. 
Taking water as the solvent, a fluid strain-rate of
$\gammadot=1~s^{-1}$ and a fiber of $L=4\, \mu m$ we find $Pe \sim 400$.
For fibers of a length greater than a few microns $Pe\gg1$ and center of
mass diffusion can thus always be neglected compared to advection by viscous flow.
Now for a material modulus of $Y=1~GPa$
and an aspect ratio $\epsilon=10^{-2}$ we find  $l_p/L\sim 10^5$ and
Brownian forces can be neglected over elastic forces. This is the case of
most of the synthetic fibers treated in this review. Decreasing the aspect
ratio to $\epsilon=10^{-3}$ and $Y$ by an order of magnitude reduces $l_p/L$ to $l_p/L\sim 1$ and
shape fluctuations resulting from Brownian forces will become important.
This is the case of semi-flexible polymers as for example actin filaments.

Finally, while this dimensionless parameter will appear naturally
later in the review, we introduce $\tilde\eta=8\pi\mu\gammadot
L^4/Ec$, where $c=-\ln(e\epsilon^2)$,  as the ration between viscous and
elastic forces, a control parameter
in many fiber-fluid problems.

\section{EXPERIMENTAL TECHNIQUES} 
\label{sec:ExpTechnique}

A limiting factor in experimentation of fiber-fluid interactions is
achieving good control on fibers in terms of their shape, dimensions,
and mechanical properties as well as good flow control. Even if
low Reynolds number experiments can be carried out at macroscopic
scales by using highly viscous suspending fluids, it is much more
comfortable and advantageous to work at microscopic scales where
microscopy and microfluidics can be combined to perform reliable
experiments. The microscopic approach requires microfabrication
techniques, smart choice of the microchannel geometry, and ways to
track the particle and its deformation while transported. All these
aspects will be detailed in the following discussion.

\subsection{Fiber synthesis and properties}
\label{subsubsec:fiberfabrication}

Reliable experiments require fibers that have controlled dimensions
and high aspect ratios as well as known and reproducible mechanical
properties. Recent fabrication techniques that can be directly
implemented into the channels are specially well-suited for this goal.


\subsubsection{Fiber fabrication}

\begin{figure}
		\includegraphics[width=\textwidth]{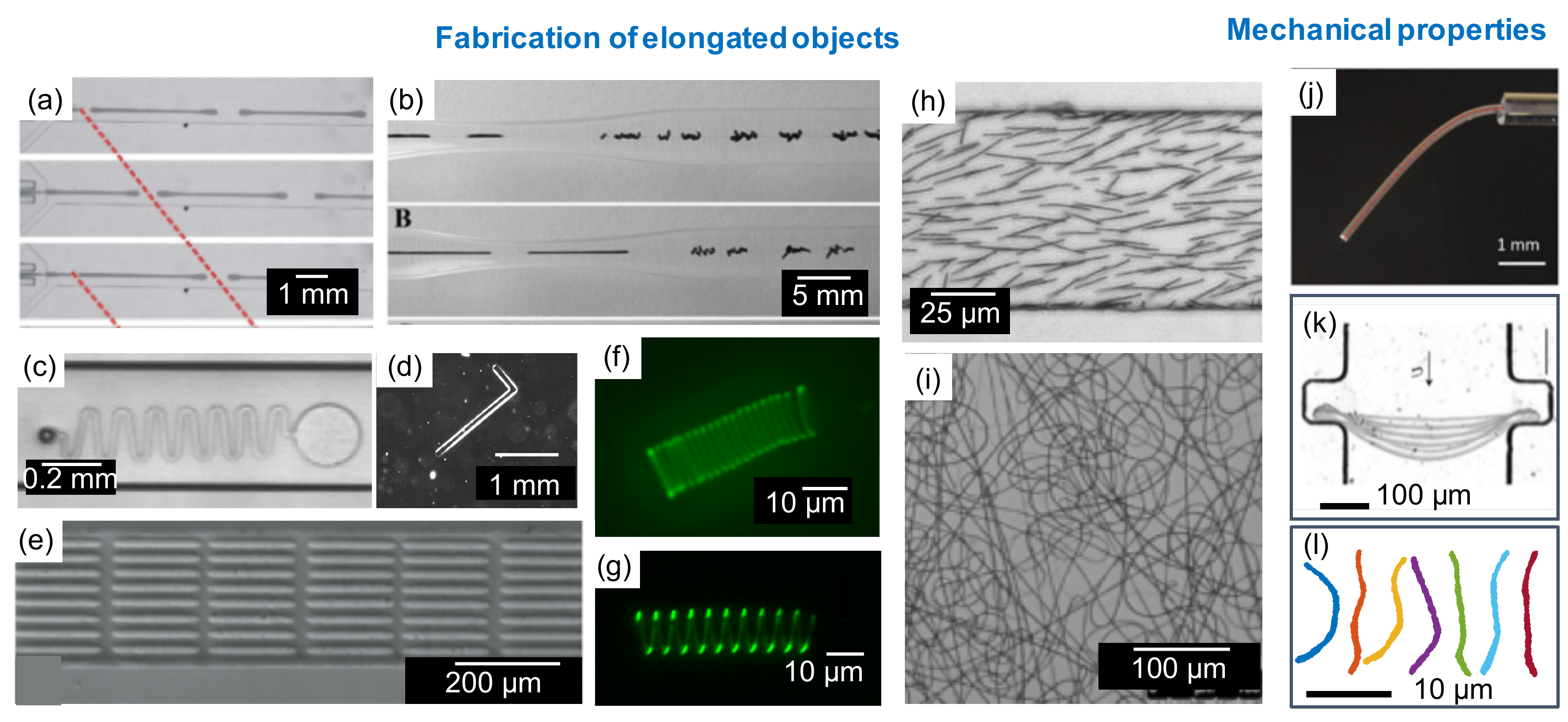}
		\caption{{\bf Fibers, their fabrication, and their mechanical properties.} {\bf(a) to (e)} Objects directly fabricated in microfluidic channels: jets
			are polymerized either by light (a) \cite{Nunes2012} or by chemicals (b) \cite{Mercader2010}. {\bf (c) to (e)} particles synthesized in a microchannel by photopolymerization of photosensitive solution through a mask. (c) \citep{Attia2009}, (d) credit: J. Cappello and   M. Da\"{\i}eff, (e) \cite{Berthet2016} {\bf(f)}  Flexible helix formed spontaneously from asymmetric ribbon, credit: M. Da\"{\i}eff. {\bf(g)} Microprinted helix, credit: F. Tesser and J. Laurent. {\bf(h)} Magnetic fiber suspension fabricated by self-assembly of magnetic colloids, \citep{Berthet2016}. {\bf(i)} Electrospun fibers, \citep{Nakielski2015}. {\bf(j) to (l)} Characterization of mechanical properties. (j) Bending of a macroscopic fiber under its own weight, \citep{Quennouz2013}. (k) Superimposed shapes of a photopolymerized fiber deformed by increasing flow rates, \citep{Duprat2015}. (l) Bending fluctuations of actin filament, credit: Y. Liu.}
		\label{fig:fibers}
\end{figure}

The fabrication of microparticles with non-spherical shapes has received much attention in recent years, but elongated objects require specific approaches. 
Electro-spinning can be an efficient technique to
fabricate fibers of typical diameters $\sim 200~nm$ and lengths of
$1-100~\mu m$ (Fig.~\ref{fig:fibers}i) with controlled mechanical
properties (\cite{Li2004a, Jun2014, Nakielski2015}). 
Microfluidics has opened new routes to tailor elongated objects. Using
the control over channel geometry given by soft lithography, it is
possible to form a jet of a solution that will solidify either by
photo-polymerization or by diffusion of small molecules from the
surrounding fluid (Figs.~\ref{fig:fibers}a\&b) (\cite{Jeong2004,
	Mercader2010, Choi2011, Nunes2012, Perazzo2017}). Another option is
to photo-polymerize a photo-sensitive solution, directly in the
microchannel, through a mask which determines the object shape
(Figs.~\ref{fig:fibers}c-e). The presence of oxygen close to the
PDMS walls prevents polymerization in these regions keeping the
fabricated object free to flow in the microchannel
(\cite{Helgeson2011, Berthet2016}). Finally,
self-assembly of colloids is a reliable technique that produces very
flexible, high aspect-ratio particles but their sedimentation due to
their high density can be limiting (Fig.~\ref{fig:fibers}h)
(\cite{Goubault2005, Zhang2011, Berthet2016}).

Another attractive strategy is to rely on biology for controlled and
monodisperse elongated objects. Actin filaments, semi-flexible
polymers resulting from the polymerization of globular protein, are models for flexible Brownian fibers (\cite{Harasim2013,
	Kirchenbuechler2014, Liu2018}). Microtubules are another candidate.
Bacterial colonies can also form elongated objects that allow
fluid-structure interaction problems to be studied (\cite{Amir2014,
	Rusconi2010, Rusconi2011}).


More complex shapes can be produced. In 2D, approaches based on
self-assembly or photo-lithography can be implemented
(Figs.~\ref{fig:fibers}c-e and \cite{Li2004a}). 3D fabrication is more
challenging, and attention has focused on microhelices given their
similarity to bacterial flagella. Electrospinning is a widespread
technology (reviewed in \cite{Silva2017}); The spontaneous formation
of helices from flat ribbons is an elegant alternative
(\cite{Pham2013}) (Fig.~\ref{fig:fibers}f). Recent developments in 3D
printing using two-photon excitation allows sub-micron resolution
but is currently limited to rigid objects (Fig.~\ref{fig:fibers}g).

\subsubsection{Mechanical Characterization}

The mechanical properties of macro-scale fibers can be measured by
rheometry or by the bending of the fiber under its own weight (see
Fig.~\ref{fig:fibers}j). When working with elastomers these two
measurements are in good agreement (\cite{Quennouz2015}). Beam-bending
measurements can be implemented in a microfluidic channel to perform
{\it in situ} measurements (see fig.\ref{fig:fibers}k)
(\cite{Duprat2015}).

When the smallest dimension of the fiber is a micron or less, Brownian
fluctuations can be large enough to induce bending deformations. In
this case the flexural rigidity  $E$ can be expressed as $l_p k_b
T$, with $l_p$ the persistence length, and an elegant way to
measure $E$ is to extract $l_p$ from the analysis of many
filament configurations (fig.~\ref{fig:fibers}l). This approach is
common for biological semi-flexible polymers like actin or
microtubules (\cite{Gittes1993}) and has also been used for tiny
electrospun filaments (\cite{Nakielski2015}).

\subsection{Flow control}
\label{subsec:flow}

\begin{figure}
		\includegraphics[width=\textwidth]{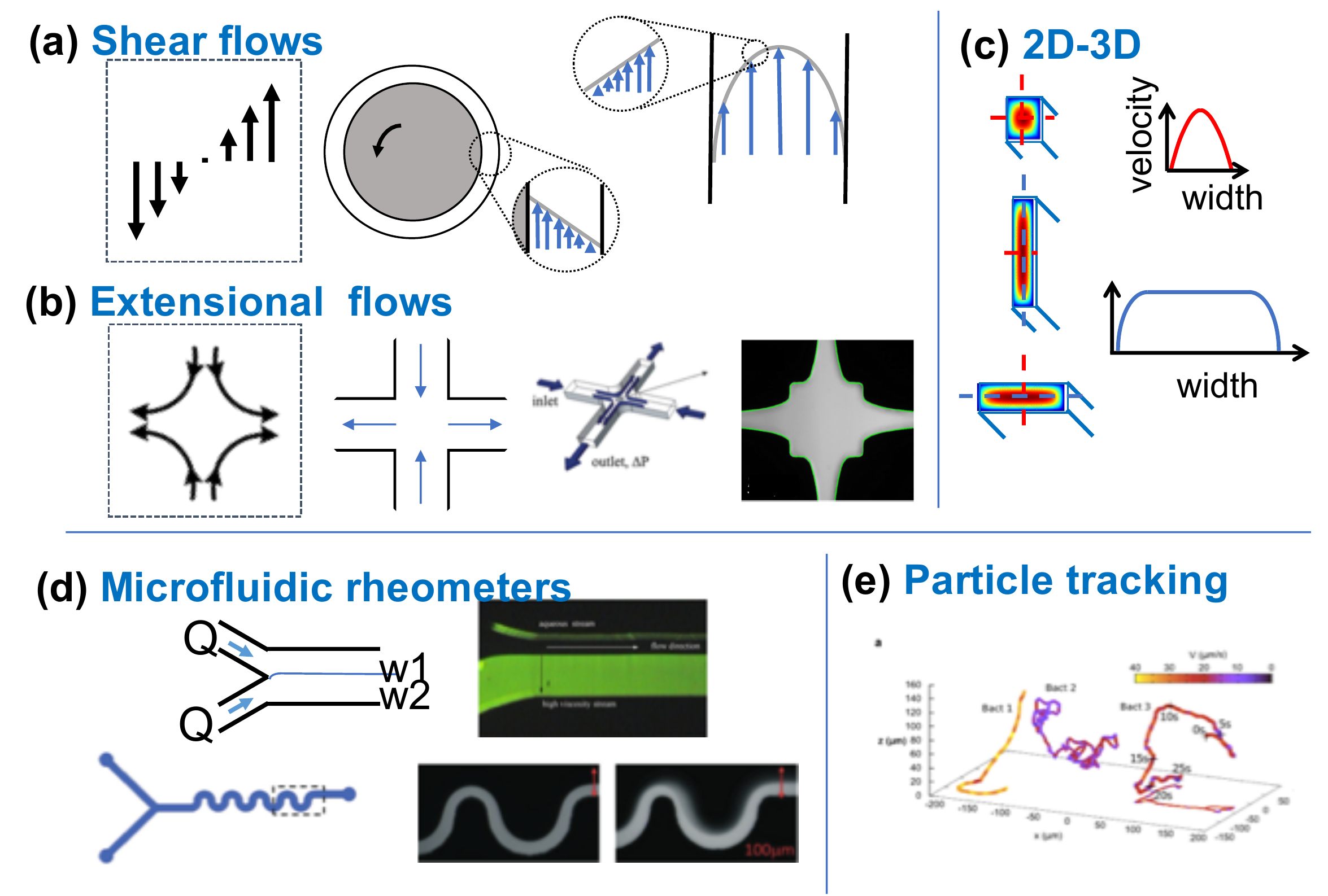}
		\caption{{\bf Flow control and fiber tracking. }Experimental realization of (a) shear flow, and (b)
			extensional flow. {\bf  (a) }From left to right, velocity profile of pure
			shear, schemes of a Couette flow, and Poiseuille flow in a channel
			which locally approximate pure shear. {\bf (b)} Velocity profile of an
			extensional flow, scheme of a cross-slot channel, experimental
			realization, \citep{KG2012}, optimized cross-slot
			geometry giving a larger area with constant extension rate, \citep{Haward2012}. {\bf (c)} Flow geometry in 3D. From top to bottom,
			square channel, vertical and horizontal Hele-Shaw cell. Velocity
			profiles corresponding to (top, red) a Poiseuille flow and (right,
			blue) a plug flow. {\bf (d)} Microfluidic rheometers. Top row, Y-channel is
			well-suited to measure small variation of viscosity, \citep{Galambos1998}. Bottom row, serpentine channels can be used to study  elastic turbulence in non-Newtonian fluids, \citep{Zilz2014}. {\bf (e)} 3D  Lagrangian tracking of bacteria in a microchannel, \citep{Darnige2017}.}
		\label{fig:flows}
\end{figure}

At large spatial scales, working at low Reynolds number requires
long-time experiments with low-velocity flows of highly viscous
liquids. However, at small scales low Reynolds numbers are easily
achievable even with low viscosity fluids. For this regime the channel
geometry determines completely the flow and thus different flows can
be obtained depending on the device design.


To avoid flow complexity, studies are often done using linear
extensional or shear flows (Fig.~\ref{fig:flows}a\&b). Experimentally,
working with linear shear flow is only possible in a Couette
geometry if the gap is small compared to the cylinder diameters
(Fig.~\ref{fig:flows}a). In channel flows, pure shear can be
approximated by Poiseuille flow if the object size is much smaller
than the channel height (Fig.~\ref{fig:flows}a)
(\cite{Liu2018}). Cross-slot geometries generate extensional flows but
with small spatial extension.  An elegant but difficult way of
overcoming this limitation is to keep the object at the stagnation
point by adjusting the pressures in the different entries as was
implemented by \cite{Schroeder2003} and later by \cite{KG2012}
(see Fig.~\ref{fig:flows}b). A more robust way is to optimize
numerically the exact geometry to obtain a purely extensional flow
over a larger distance even in the presence of the walls
(Fig.\ref{fig:flows}b) (\cite{Haward2012, Zografos2016}).

Numerical simulations are sometimes done in 2D due to their relative
simplicity, while in experiments 2D flows are not always easy to
achieve. One option is to use a Hele-Shaw geometry, in which
lateral dimensions are much larger than the cross-channel dimension
(Fig.~\ref{fig:flows}c). On a microscope,
the vertical configuration allows a nearly 2D flow to be generated in the
observation plane above the boundary (\cite{Haward2012}).

Flows can be controlled by syringe pumps imposing flow rates, or by
pressure control devices. Pressure control is well adapted to low
velocity and shear rate. Particle velocimetry techniques are used to
quantitatively determine local velocities and thus shear rates.

A new development in the field are microfluidic rheometers that access
the rheological properties of complex suspensions of various types
(\cite{Lindner2016}) (Fig.~\ref{fig:flows}d). Their main advantages
are their sensitivity to low variations in viscosity and normal
stresses and the possibility to directly observe the flowing particles
within the device.

\subsection{Particle observations and tracking}

Typically, fiber deformations and trajectories need to be measured
simultaneously even though they occur at different spatial scales:
Images with high-magnification are necessary to obtain good resolution
of fiber deformations whereas trajectories are inherently large-scale.
In a microchannel, bright field, phase contrast, and fluorescence
microscopy are usually best suited to acquire sufficient quality
images of fiber deformations. The use of high-magnification and a
high-numerical-aperture objective is in general required. In addition, a
long working distance can be very useful to observe filaments far from
the bottom of the chamber where the coupling to the wall is
negligible. At macro-scale, the 3D-shape may be reconstructed from
images taken by two cameras or by the smart use of mirrors.

To track an object along its trajectory at the macro-scale, automated
motion of the camera can be used. Keeping a fiber in the field of view
in a microfluidic channel requires that the microscope stage moves at
or near the fiber velocity. In combination with a fast camera this
avoids image blurring.  Following trajectories in 3D becomes necessary
when objects do not follow streamlines as in cross-stream migration,
or for active particles. In this case, one can use 3D tracking
algorithms like those developed to follow swimming bacteria
(\cite{Darnige2017, Qu2018}).

\section{MATHEMATICAL MODELING AND SIMULATION METHODS} 
\label{sec:MathMod}

The interaction of flexible fibers with Stokesian fluids
has a special structure for which specialized mathematical and
computational methods have been developed. The most basic is local
slender-body theory (SBT), which gives a leading-order local relation
between elastic and drag forces and is a mainstay for analysis.
Nonlocal interactions arising from fluid incompressibility can be
captured through use of higher-order, and more complex, slender-body
formulations, or through other approaches such as immersed boundary
methods, bead-rod models, or regularized Stokeslet methods. Here we
first discuss the nonlocal SBT formulation, as it is compact and amenable
to fast solution methodologies.

\subsection{Nonlocal Slender Body Theory}
\label{NSBT}

The appeal of nonlocal SBT is its reduction of filament-fluid and
filament-filament interactions to relatively simple dynamics equations
for filament centerlines. For this we assume that the fluid is
described by the incompressible Newtonian Stokes equation:
\begin{equation}
\label{eqn:stokes}
\mu \Delta \bu -\nabla p = \bzero \mbox{~~\&~~}\nabla\cdot\bu=0,
\end{equation}
with $\mu$ the shear viscosity, $\mathbf{u}$ the
fluid velocity, and $p$ the pressure. Generally, solutions to the
Stokes equations can be found using boundary integral methods
(\cite{Pozrikidis1992}), where the fluid velocity is represented as a
distribution of fundamental solutions to the Stokes equations on all
immersed and bounding surfaces. The fundamental solution
corresponding to a $\mathbf{\delta}$-function of force is given by the
Stokeslet tensor: $\mathbf{G}(\bx)=\frac{1}{8\pi\mu}
\frac{\mathbf{I}+\hat\bx\hat\bx}
{|\bx|}$, where $\hat\bx=\bx/|\bx|$.

The flows produced by the motion of slender fibers are treated
especially as their surface integrals can be reduced, through
asymptotics, to integrals of Stokeslets along their center-lines
(\cite{KR1976,Johnson1980,Gotz2000}). Specifically, consider $N$ fibers,
each of radius $a$ and lengths $L_n$, with center-line positions
$\bX^n(s,t)$ for $n=1, ..., N$, with $s$ the arclength ($0\leq s \leq
L_n$).  Each fiber is assumed to be slender, i.e.,
$\epsilon_n=a/L_n\ll 1$, and to exert a force per unit length
$\bff^n(s,t)$ on the surrounding fluid. Given a background flow
$\barbu$ the fluid velocity at a point $\bx$ within the fluid is, to
leading order, given by distributions of Stokeslets along
fiber center-lines:
\begin{align}
\mathbf{u}(\mathbf{x})=\bar{\mathbf{u}}(\mathbf{x})+\sum_{m=1}^{N} \mathbf{u}_F^m(\mathbf{x})
\mbox{~~~~where~~~~} 
& \mathbf{u}^m_F(\mathbf{x}) 
= \int_0^{L_m}\bG(\bx-\bX^m(s'))\cdot
\bff^m (s')ds'
\label{FluidVeloc}
\end{align}

\cite{Gotz2000} derives higher-order corrections in terms of
integrals of Stokes doublets.

The self-induced velocity of a fiber is not found by direct evaluation
of Eqs.~(\ref{FluidVeloc}) upon that fiber. This yields a
logarithmically divergent integral. Instead, fiber self-induction is a
matter of careful asymptotic matching
(\cite{KR1976,Johnson1980,Gotz2000}). Such analyses give the velocity of
the centerline $\bX^n(s)$ as
\begin{equation}
\bV^n(s)=\barbu(\bX^n)
+\sum_{m=1, m\neq n}^{N}\mathbf{u}_F^m(\mathbf{X}^n)
+c_m\left(\bI+\bX^n_s\bX^n_s \right)\bff^m
+\bK_n[\bff^n]
\label{NonlocalSBT}
\end{equation}
where $\mathbf{X}_s$ is the tangent vector and subscript $s$
denotes a partial derivative. Here $c_n=-\ln(\epsilon_n e^2)$ is an
asymptotically large parameter, and $\bK_n$ is an $O(\epsilon^0)$
linear operator that contains a "finite-part" integral of Stokeslets
that captures fiber self-interactions. The use of this integral in
computation is discussed by \cite{TS2004} who show that it
must be appropriately regularized both to maintain asymptotic accuracy
and to control the growth of short wavelength modes treated
inaccurately by slender-body theory.

Equation (\ref{NonlocalSBT}) relates the fiber velocity to the fiber
forces acting upon the fluid.  As a dynamics problem, this system is
closed by specifying the fiber elastic forces and identifying a
Lagrangian or material variable for the fiber. These are done
simultaneously by modeling each fiber as an inextensible
Euler-Bernoulli beam:
\begin{equation}
\bff^n=\bff^{n}_{\text{E}}=-E\bX^n_{ssss}+(T^n\bX^n_s)_s,~~0\leq s \leq L_n~.
\label{EBE}
\end{equation}
with $-E\bX_{ssss}$ the bending force per unit length, and $T$ the
axial tension with $(T\mathbf{X}_{s})_s$ the tension force per unit
length. Inextensibility identifies $s$ as a material variable, and
hence $\bX_t^n(s,t)=\bV^n$. Inextensibility also generates an
auxiliary integro-differential equation for $T$ by imposing
$\mathbf{V}^n_{s} \cdot \bX^n_{s}=0$ which follows from
differentiating the identity $\bX^n_s\cdot\bX^n_s=1$ (\cite{TS2004}).

\subsection{The special case of local SBT}
\label{sec:SBT}

Local SBT has been the mainstay for theoretical analyses of the
dynamics of flexible fibers, and sits at the core of many numerical
approaches. In particular, retaining from Eq.~(\ref{NonlocalSBT}) only
the logarithmic leading order in $\epsilon$ we have a local
balance of fluid drag forces, and fiber forces exerted on the fluid:
\begin{equation}
\bR(s,t)\cdot \left[\bV(s,t)-\barbu(\bX(s,t),t)\right]
=\bff(s,t),
\label{LocalSBT}
\end{equation}
where
$\bR=\eta \left(\bI-\mathbf{X}_s\mathbf{X}_s/2\right)$ and $\eta=8 \pi \mu/c$.
For an elastic fiber modeled by Eq.~(\ref{EBE})) we have a
dynamical equation, plus a constraint:
\begin{equation}
\bX_t= \bV \equiv \bu(\bX,t)+\eta^{-1}(\bI+\bX_s\bX_s)
\left[-E\bX_{ssss}+(T\bX_{s})_{s}\right]~~\&~~ \bX_s\cdot\bV_s = 0
\label{LSB_EB}
\end{equation}
The inextensibility constraint generates an elliptic equation for $T$:
\begin{equation}
2T_{ss}-|\bX_{ss}|^2 T=S,
\label{tension}
\end{equation} where $S$ depends upon background flow and
the bending force.  Given appropriate boundary conditions, $T$ has a
unique solution, and Eq.~(\ref{LSB_EB}) can be used to evolve the
fiber's shape, position, and orientation.

{\it Basic scaling}: Consider a background flow with a characteristic
length-scale $W$ and time-scale $\gammadot^{-1}$, expressed as
$\barbu=\gammadot W \bU(\bx/W,\gammadot t)$. By scaling space on $L$,
time on $\gammadot^{-1}$, and tension $T$ on $E/L^2$,
Eq.~(\ref{LSB_EB}) can be given in nondimensional form as
\begin{equation}
\bV = \alpha^{-1}\bU(\alpha\bX,t)
-\tilde\eta^{-1}(\bI+\bX_s\bX_s)
\left[\bX_{ssss}-(T\bX_{s})_{s}\right]
\label{ElastoFlow2}
\end{equation}
with $\alpha=L/W$, and where $\tilde\eta=8\pi\mu\gammadot L^4/Ec$ is
the effective strength of flow forcing. Note that for linear
background flows $\alpha$ disappears from the equation. If the tension
is negative, reflecting compressive fluid stresses, then in
Eq.~(\ref{ElastoFlow2}) is seen as a competition of fourth-order
diffusion and a second-order anti-diffusion.

\subsubsection{Boundary conditions} 
Finally, boundary conditions are needed on the tensile and bending
forces. For freely suspended fibers, these are the ``natural''
boundary conditions of having zero applied forces and torques, which
simplify to $\mathbf{X}_{sss}=\mathbf{X}_{ss}=\mathbf{0}$, and $T=0$
on the each end of the fiber. These boundary conditions suit the fiber
force containing $\bX_{ssss}$ and $T$ satisfying essentially an
elliptic boundary value problem. Having filaments attached (clamped or
hinged) to surfaces, mobile or immobile, generates other boundary
conditions and explores other dynamics, which are detailed in
\cite{NRZS2017}.

\subsection{Brownian semi-flexible fibers}
\label{sec:BLSB}

Thus far we have assumed the fibers to be rigid enough that
thermal fluctuations have negligible effect upon their
conformations. For this to hold, the length of the fibers needs to be
significantly smaller than their persistence length $l_p$.
If this assumption is not met (as can be for carbon nanotubes, actin
filaments, and wormlike polymers), thermal fluctuations can have
important consequences on the overall mechanics of the
suspension, and has been the subject of many studies.  A review of
this topic is beyond the scope here and we refer the
interested reader to \cite{Broedersz2014}. Here we only give a
limited summary of previous work and the standing
challenges.

In the presence of thermal fluctuations Eq.~(\ref{LocalSBT})
is modified to
\begin{equation}
\bR\cdot \left(\mathbf{X}_t-\barbu \right)=-E\mathbf{X}_{ssss}+\left(T\mathbf{X}_s\right)_s +\mathbf{f}^{B}(s)
\label{LSB_B}
\end{equation}
where $\mathbf{f}^{B}$ is the thermal fluctuating force with zero mean
and finite variance given by fluctuation dissipation theorem:
\begin{equation}
\langle \mathbf{f}^{B}(s,t)\rangle =\b0~~\&~~
\langle \mathbf{f}^{B}(s,t) \mathbf{f}^{B}(s^\prime,t^\prime)\rangle
=2kT \bR(s,t)\delta (s-s^\prime)\delta (t-t^\prime)
\label{FDT}
\end{equation}
The brackets denote ensemble averages.  \cite{Munk2006} used this
formulation to study a 2D Brownian semi-flexible fiber in shear flow.
\cite{Manikantan2013} also used this formulation in studies of
Brownian flexible fibers in two-dimensional arrays of counter-rotating
Taylor-Green vortices and extensional flows.

It follows from Eq.~(\ref{FDT}) that computing the displacements from
thermal fluctuations requires computing $\bR^{1/2}$. For local SBT
$\bR^{1/2}$ can be computed analytically. For non-local SBT, the
resistance tensor, $\bR$, is nonlocal and depends on the position of
all fibers.  Numerically, this means computing the square root of a
square matrix with dimensions $N_T^2=(N\times N_F)^2$, where $N_F$ is
the number of fibers and $N$ is the number of discretization points
per fiber. The direct computational of $\bR^{1/2}$ requires
$\mathcal{O}(N_T^3)$ operations, which makes simulation of systems
beyond a few fibers very expensive. More efficient techniques have
been proposed to reduce the computational cost to $N_T^\alpha$ with
$\alpha$ varying between $1<\alpha<2$, depending on the conditions. In
a recent development \cite{Fiore2017} use Ewald summation to reduce
the computational cost to $O(N_T)$.

Several other works have modeled Brownian filaments by using
bead-rod/spring models for inextensible polymeric filaments. The main
complexity involves imposing the inextensibility constraint.
Specifically, it turns out that very stiff springs and inextensible
rods differ in their degrees of freedom and hence in the resulting
equilibrium statistical mechanics \citep{Fixman1978}.
\cite{Hinch1994} showed that a psuedo-potential force needs be applied
to a connection with a very stiff spring in order to exactly recover
the behavior and statistical physics of an inextensible connection.

\subsection{Fast Numerical Methods for SBT}
\label{sec:FastSBT}

\cite{TS2004} developed a stable and numerically tractable version of
nonlocal SBT for flexible filaments with free-ends, and devised
specialized quadrature schemes for the regularized finite-part
integral in $\bK[\bff]$ of Eq.~(\ref{NonlocalSBT}). The fibers'
centerlines were discretized uniformly and derivatives evaluated using
a second-order accurate finite difference scheme, while trapezoidal
rule was used for evaluating the remaining integrals. For time-stepping, a
backward differentiation finite difference scheme was used and the
bending forces were treated implicitly to remove temporal stiffness,
while the tension equation ? which imposed inextensibility of the
fibers ? was treated explicitly.

\cite{NRZS2017} extended the work of \cite{TS2004} to a
platform suitable for simulating a substantially larger number
of fiber, to $O(1000)$ or more. They used a Chebyshev
basis for the representation of fiber position and to compute
high-order derivatives with spectral accuracy. They used
implicit-explicit time discretization and treated both bending and
tensile forces implicitly. As a result, they observed a three order of
magnitude improvement in the maximum stable time-step in their
numerical experiments, compared to the explicit treatment of tension
by Tornberg \& Shelley. They solved the linearized system of equations
using GMRES (\cite{Saad1986}) with a Jacobi or block Jacobi
preconditioner. A kernel independent fast multipole method (FMM) 
is
utilized for fast computation of nonlocal hydrodynamic interactions
via fast matrix?vector products, yielding a computational cost of
$O(M)$ operations per time-step, where $M$ is the number of
unknowns. The entire computational scheme is parallelized and scalable
to many computational cores, which allowed simulation of $O(1000)$
semi-flexible filaments.

\subsection{The Regularized Stokeslet Method}

A popular method for solving the Stokes equations is the method of
regularized Stokeslets of \cite{Cortez2001} which uses superpositions
of regularized fundamental solutions.  \cite{FLMTC2005} used a
superposition of regularized Stokeslets and Rotlets to simulate the
dynamics of driven flagellae. In their study, a flagellum is a network
of flexible springs, and a helical shape so composed is driven by a
torque at its base. In a study also modeling flagella (and cilia)
\cite{Smith2009} developed a version of regularized Stokeslets that
utilizes a boundary-element approach to the discretization and
incorporates the presence of walls (see also \cite{JWYZ2014}).
\cite{BLY2011} use a regularized representation of a one-dimensional
curve of two-dimensional Stokeslets to simulate the nonlocal dynamics
of flexible, slightly extensible fibers. \cite{OLC2013} have combined
the regularized Stokeslet method for evolving slender rods with the
internal mechanics of an elastica with intrinsic twist and curvature.
And, in theoretical work, \cite{CN2012} have developed a
asymptotically consistent slender-body theory based on using
regularized center-line forces.

\subsection{The Immersed Boundary Method}
The immersed boundary method (IBM) of \citep{Peskin2002} has also been
applied to this class of problems. In this method, a filament is
discretized by Lagrangian markers connected by spring elements, and
their relative displacements by fluid motions are used to calculate
the filament's elastic response. These elastic forces are then
distributed onto a background grid covering the fluid volume, and are
used as forces acting upon the fluid, thus modifying the fluid flow
that displaces the markers \citep{Lim2004}. The advantage of the IBM
is that detailed immersed mechanical structures can be simulated, but
at the cost of having to solve the flow equations in the entire fluid
volume. Other volume-based methods have also been used in this
context, including finite element/volume methods \citep{Mitran2007},
and Lattice Boltzmann \citep{Wu2010}.

Within the immersed boundary framework, for a fiber to have a
physical width the volume grid size needs to be several times
smaller, making the method computationally expensive
for simulating slender bodies such as fibers and thin disks.
To overcome this, \cite{NF2014} employed an adaptive-mesh version of
the IBM \citep{GHMP2007} to study the dynamics of
flexible fibers as models, in part, for diatom chains (diatoms are
unicellular phytoplankton). For this, they constructed fibers that are
composite structures made of alternating segments that mimic the
structure of diatom chains (see Fig.~\ref{fig:Shear}c).  While a simple
beam model, like Eq.~(\ref{ElastoFlow2}), describes its bending
deformations well, the results of compressive strain are
not. \cite{Wiens2016} used their parallel implementation of the
IBM to simulate a suspension of up to 256
semi-flexible fibers in shear flow. Their analysis of parallel
performance shows their algorithm is weakly scalable. Recently,
\cite{Stein2017} implemented a fourth-order accurate IBM
for studying flows over arbitrary smooth domains. A combination
of this with adaptive-grid immersed boundary \citep{GHMP2007} might
improve the computational efficiency and accuracy of immersed boundary
in simulating slender objects in flows.

\subsection{Bead-and-Rod Models}
While closely related to methods based upon SBT or the immersed
boundary method, bead-and-rod models have a somewhat independent
lineage. In bead-and-rod models, a flexible fiber is represented as a
one-dimensional chain of linked rigid bodies (e.g. spheres, spheroids,
rods) that experience local drag and interact with each other through
short-range forces (e.g. repulsion, lubrication, friction), while
sometimes neglecting long-ranged hydrodynamic interactions or only
including a subset of them. A recent review of these methods is found
in \cite{HLHN2011}. In recent work \cite{Delmotte2015} have elaborated
upon basic bead models by introducing a new Lagrange multiplier method
to impose constraints, and consider several flow problems -- Jeffery
orbits, buckling in shear, actuated swimming filaments -- by using an
approximate accounting of Stokesian hydrodynamics.

\section{FREE FIBERS IN FLOW}
\label{sec:FreelyTransportedFibers}

How fibers are buckled by flow is central to much of the interesting
nonlinear dynamics observed in simulations and experiments of fiber
motion. Here we first discuss a prototypical situation -- a straight
fiber moving in a linear straining background flow -- where buckling
arises as an instability. We then discuss other prototypical problems
such as morphological transitions of fibers transported in linear
shear flows and more complex flow geometries where the coupling of
fiber deformation and transport can be observed.

\subsection{Fiber morphology in simple flows}

Here we discuss the shape dynamics of fibers in linear shearing and straining flows. 

\subsubsection{Buckling instabilities in compressive flows}
\label{subsubsec:buckling}

\begin{figure*}
		\includegraphics[width=\textwidth]{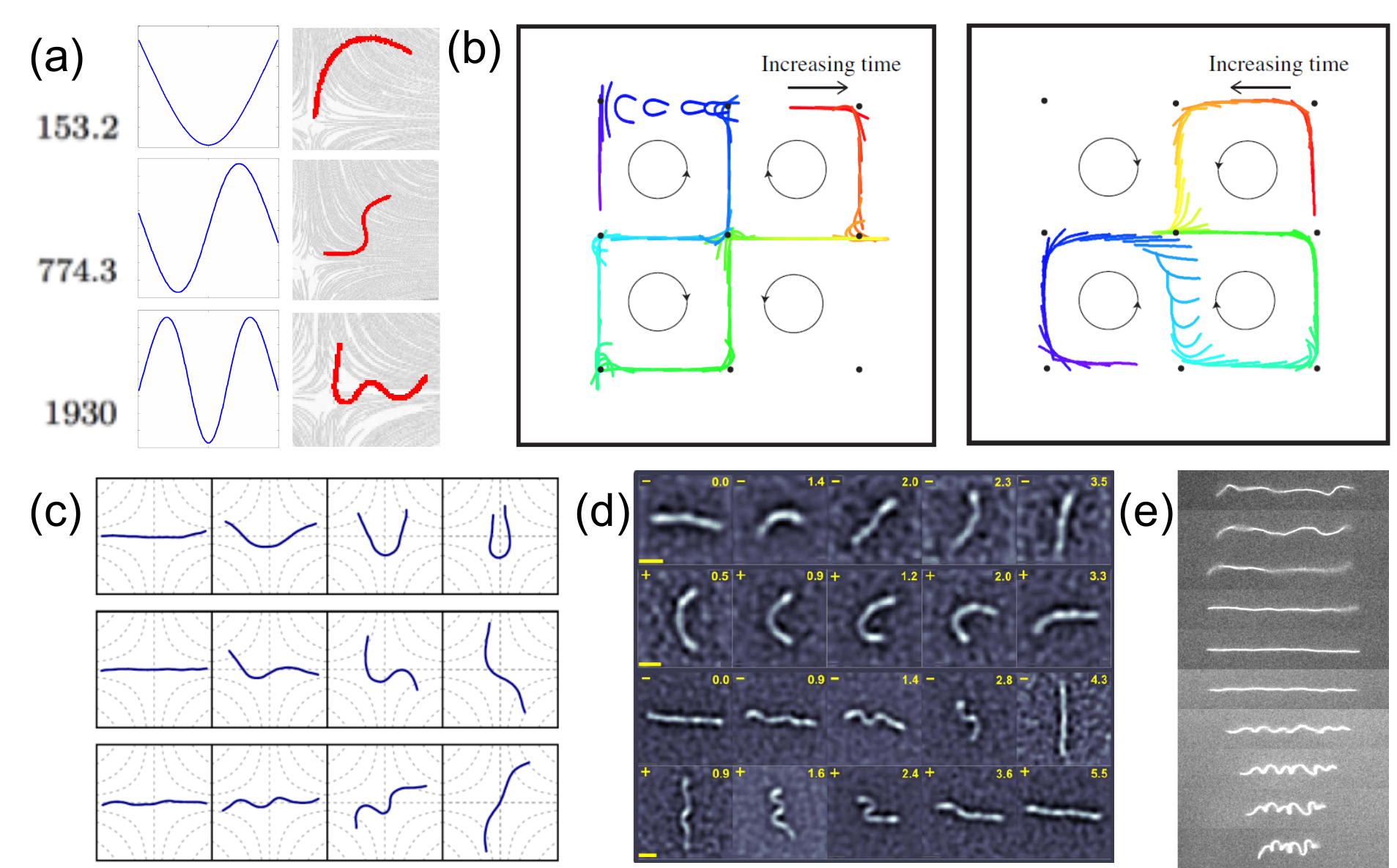}
		\caption
		{{\bf Fiber buckling.}  {\bf (a)} Critical values of the control
			parameter $\tilde\eta$ and fiber shapes of
			buckling modes from linear stability analysis and experiments, \citep{Quennouz2015}. {\bf (b)} Flexible fiber evolving in an array of hyperbolic stagnation
			points, $\tilde\eta\approx2000$, \citep{Quennouz2015}: Left,
			simulations using SBT, right, experiments with a polymeric centimetric fiber.  {\bf (c) }\&{ \bf (d)}
			Brownian fiber maintained at a stagnation point. $\tilde\eta$ increases from top to bottom. (c)
			Numerical simulations using SBT, \citep{Manikantan2015}. (d) Experiments using actin
			filaments in a microfluidic device, \citep{KG2012}.  {\bf (e)} Actin filament
			transported in a converging/diverging hyperbolic
			microchannel experiencing
			elongation followed by compression at $\tilde\eta=7.9 \, 10^5$, credit: Y. Liu.}
		\label{fig:buckling}
\end{figure*}

Because of its simplicity, local SBT is usually the preferred tool for
studying linear stability of immersed fibers and is used here to
illustrate the buckling instability of free fibers in the simplest of
cases.

A perfectly straight isolated fiber will remain straight in any
incompressible and linear flow, even as the fiber is translated and
rotated. Hence, consider the background flow
$\bU(\bx,t)=\bA(t)\cdot\bx$ with $tr(\bA)=0$. A straight fiber can be
represented as $\bX(s,t)=\bX_c(t)+s\be(t)$ with $\bX_c$ its center
point, and $\be$ a unit orientation vector. Inserting this form into
Eqs.~(\ref{ElastoFlow2}) \& (\ref{tension}), and applying the
``natural'' conditions of zero total force and torque, yields
\begin{equation}
{\dot\bX}_c=\bA\bX_c,~~{\dot\be}=\left(\bI-\be\be^T \right)\bA\be,
~~{\bar T}=-\frac{\tilde\eta}{4}(\be^T\bE\be)(s^2-1/4)
\label{StraightRod}
\end{equation}
where $\bE=(\bA+\bA^T)/2$ is the symmetric rate-of-strain
tensor.  Hence, the rod center is carried with the local flow, the
orientation vector obeys Jeffrey's equation (\cite{Jeffrey1922}), and
the tension ${\bar T}$ is quadratic in $s^2$ with its sign
determined by the orientation of $\be$ relative to the principle axes
of $\bE$. Thus, if the fiber is aligned with compressive straining of
the flow then the tension is negative and hence compressive. This is
the condition required for a buckling instability.

The most straightforward case is given by a simple straining flow
$\barbu=(x,-y)$ where the straight fiber is aligned with the
$y$-axis (i.e., $\be\equiv{\hat\by}$). This choice maximizes flow
compression (\cite{YS2007}) and is free of explicit time-dependence from
the background flow. This leads to a variable coefficient eigenvalue
problem for the growth or decay of an in-plane perturbation $w$ to the
straight fiber:
\begin{equation}
\sigma w=w+sw_{s}+\frac{1}{4}\left(s^{2}-1/4\right)w_{ss}-\tilde\eta^{-1}w_{ssss}
\label{LinearEq2}
\end{equation}
where $w_{ss}=w_{sss}=0$ at $s=\pm1/2$.  With an eigenvalue solver we
can track the system's eigenvalues and eigenfunctions as $\tilde\eta$,
the effective viscosity or strain-rate, is increased. For small
$\tilde\eta$ the straight fiber is stable to perturbations.  With
increase in $\tilde\eta$ we find the successive crossing to the right
half-plane of eigenvalues coupled to eigenfunctions associated with
increasingly higher order bending modes. The first three crossings
occur at $\tilde\eta_1=153.2$, $\tilde\eta_2=774.3,$ and
$\tilde\eta_3=1930$ (\cite{Quennouz2015}), and their associated
eigenfunctions are shown in Fig.~\ref{fig:buckling}a (left panel). The
respective eigen-shapes are the classical U-, S-, and W-shaped
buckling modes.

In experiment it is nontrivial to achieve a linear straining flow and
reasonably long residence times for a fiber moving in it. One
possibility is to use the approach of a fiber to a stagnation
point. Fiber buckling has been investigated in a macroscopic
realization of such a system by \cite{WQFLdR2010} who used centimetric
soft elastomer fibers. These fibers move in a viscous quasi-2D
cellular flow, an array of counter-rotating vortices surrounding hyperbolic stagnation
points. Above a critical value of $\tilde\eta$,
fibers are observed to buckle (Fig. \ref{fig:buckling}b (right)) at or
near stagnation points.  With increasing $\tilde\eta$ more complex
fiber shapes were observed  (Fig.~\ref{fig:buckling}a (right)), corresponding to the eigen-shapes
predicted by the linear stability analysis. Fiber trajectories from experiments and simulations
(\ref{fig:buckling}b) show fiber transport across the cellular array. Due
to the complexity of the flow, buckling does not take place at each
passage of a stagnation point, but depends on the exact conformation
and position of the approaching fiber (\cite{Quennouz2015}).

Buckling instabilities have also been investigated in the presence of
Brownian fluctuations. \cite{Manikantan2015} have shown numerically
that Brownian fluctuations play a minor role on the buckling
instability and serve mainly to broaden the threshold of the
transitions towards the different buckling modes. \cite{KG2012} investigated the deformation
of a micrometric actin fiber held at
a stagnation point created in a microfluidic cross-slot device. Their
experiments as well as numerical simulations by \cite{Manikantan2015}
show modes qualitatively similar to the deterministic predictions,
albeit with shape fluctuations owing to Brownian fluctuations (see
Figs.~\ref{fig:buckling}c, d).

The optimal flow geometry for the investigation of buckling
instabilities is a hyperbolic channel, where constant extension rates
can be achieved over long residence times. This was realized
recently using actin filaments transported in an optimized
microfluidic flow geometry (sec. \ref{subsec:flow}) and preliminary
results can be seen on Fig.~\ref{fig:buckling}e. The filament is first
stretched in the converging part of the channel where Brownian fluctuations are suppressed by the viscous forces. Then the
filament enters the diverging part of the channel where it experiences
strong compression leading to higher buckling modes. A similar
succession of stretching and coiling has been observed in
microchannels with constrictions for actin filaments
(\cite{Strelnikova2017}) and polymeric microfibers (\cite{Nunes2012}).

\subsubsection{Fiber morphology in shear flows}
\label{subsubsec:shear}

A linear shear flow is the superimposition of a rotational and a
straining flow and, as a consequence, fiber dynamics are a combination
of tumbling and periodic deformation. A rigid ellipsoid tumbles
periodically in a linear shear flow,
while resisting stretching or compression as it rotates through the
extensive and compressive quadrants. A flexible fiber on the other
hand will react to the straining flow, leading to a transition to a
buckling instability as $\tilde\eta$ is increased (\cite{BS2001,
	Liu2018, Forgacs1959a}). When $\tilde\eta$ is increased further,
fibers become too flexible to sustain rotation of the fiber as a whole
and a transition to "snaking" motion is observed. Here the fiber
remains mainly aligned with the flow direction, but performs a tank
trading motion (\cite{Forgacs1959b, Harasim2013, Liu2018,NF2014,
	Delmotte2015}).

\begin{figure}
		\includegraphics[width=\textwidth]{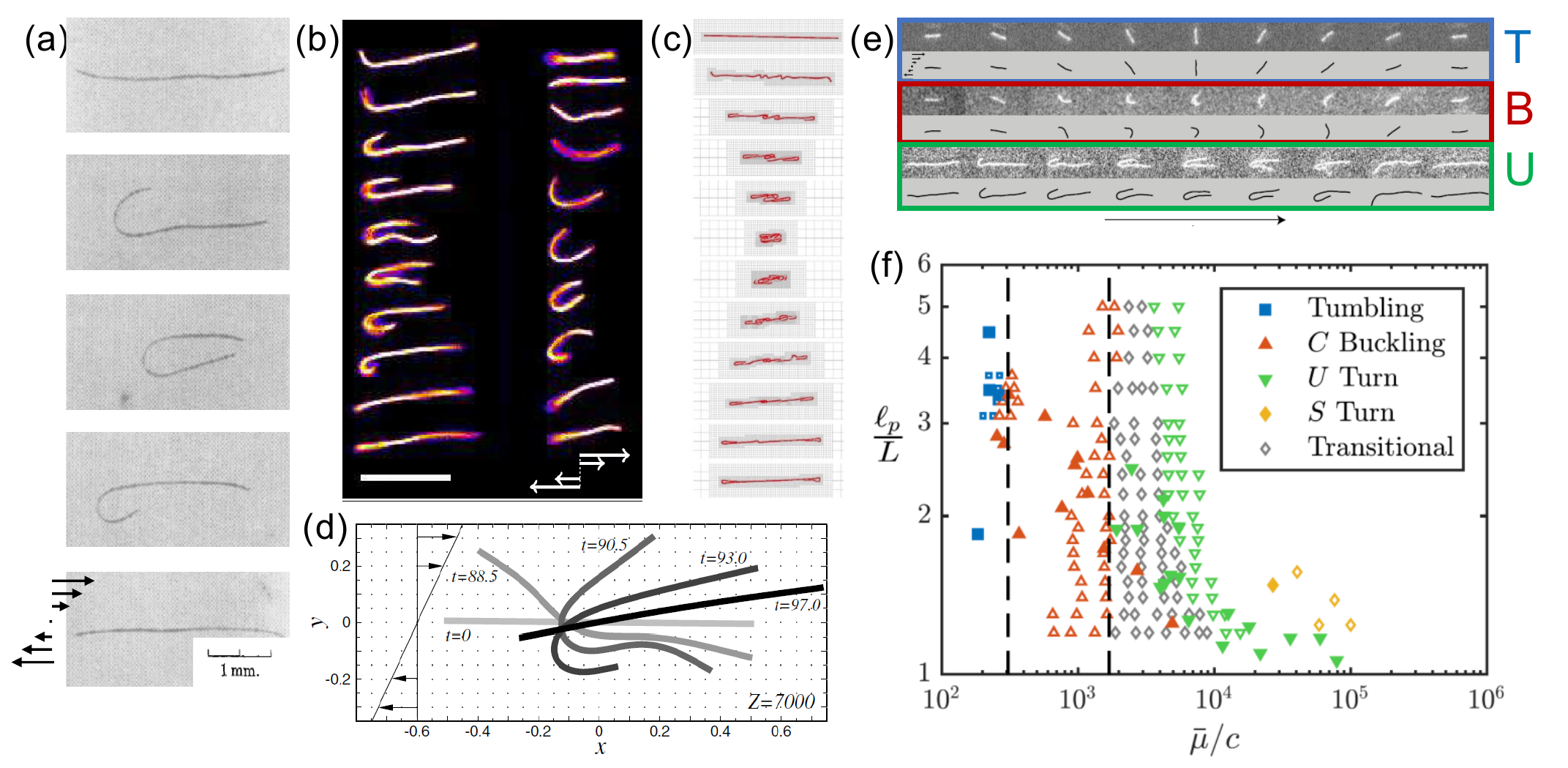}
		\caption{{\bf Fibers in shear flows.} {\bf (a)} Elastomeric fibers
			performing snake-turns in corn syrup in a simple shear
			flow, \citep{Forgacs1959b}.  {\bf (b)} Actin
			filaments in a microfluidic geometry, \citep{Harasim2013}. {\bf (c)} Complex fiber shapes from simulations
			using an adaptive version of the immersed boundary method, \citep{NF2014}.
			{\bf (d)} Dynamics of a buckling fiber at
			$\tilde\eta=7000$ from simulations using local SBT, \citep{BS2001}.
			{\bf (e)} Different morphologies of Brownian fibers with increasing length
			from experiments using actin filaments (top) and
			simulations using SBT (bottom), \citep{Liu2018}. {\bf (f)} Phase diagram on
			filament morphologies. The vertical lines indicate the transitions to buckling and snaking motions, \citep{Liu2018}.}
		\label{fig:Shear}
\end{figure}

Early experiments by \cite{Forgacs1959b} investigated the deformation
of millimetric elastic fibers in corn syrup under linear shear in a Couette cell. They were the first to address fiber buckling (\cite{Forgacs1959a})
and observed what they baptized "springy
rotations" (\cite{Forgacs1959b}). For longer fibers they identified "snake" turns
as shown on Fig.~\ref{fig:Shear}a. The transition to fiber buckling was
clearly identified only later in a numerical investigation by
\cite{BS2001} for a fiber rotating in a linear shear flow.
The first transition to buckling takes place at
$\tilde\eta=306.4$. This is twice the value $\tilde\eta_1=153.2$ found
for straining flow, as the magnitude of the straining part in shear
flow is proportional to only half of the magnitude of the shear
flow. The two threshold values are thus effectively identical.  Using
the local SBT formulation, their numerical simulations studied the
very nonlinear shape dynamics of the fiber above the buckling
transition (see Fig.~\ref{fig:Shear}d). Experiments by
\cite{Harasim2013} using actin filaments in microscopic channel flows
have investigated in detail the "snake" turn motions and were the
first to describe and model this configuration as a tank trading
motion where a narrow bend travels along the filament whose two ends
remain mainly aligned with the flow. They also addressed the role of
rotational diffusion on the period of Jeffery orbits. Numerous
investigations have observed the "snaking" dynamics numerically
(\cite{NF2014, Stockie1998, Delmotte2015}) (see
Fig.~\ref{fig:Shear}c). Only recently have all subsequent regimes been
observed (Fig.~\ref{fig:Shear}e) and classified
(Fig.~\ref{fig:Shear}f) as a function of $\tilde\eta$ and $\l_p/L$, by
experiments using actin filaments and a numerical investigation using
non-local SBT in the presence of Brownian fluctuations
(\cite{Liu2018}). Special care was taken when designing the
experimental flow geometry to assure linear shear conditions
(sec. \ref{subsec:flow}).
The authors also addressed the transition to the "snaking"
mode which is initiated by a buckling instability occurring, under sufficiently strong shear, in a filament
that has not yet aligned with the compressive axis.
Rotation of the deformed fiber will then align one end with the
flow direction whereas the other can bend. The theoretically predicted threshold is in good agreement with experimental observations (Fig.~\ref{fig:Shear}f). Well above the onset of the "snaking" motion more complex dynamics
such as coiling (\cite{NF2014}), knot formation (\cite{Kuei2015}), or
helical motion (\cite{Forgacs1959b}) were also
reported.

\subsection{Fiber morphology and transport in more complex flows}

Here we discuss the shape dynamics of fibers in more complex background flows, such as those arising from confinement or imposition of a background flow through external forcing. 

\begin{figure}
		\includegraphics[width=\textwidth]{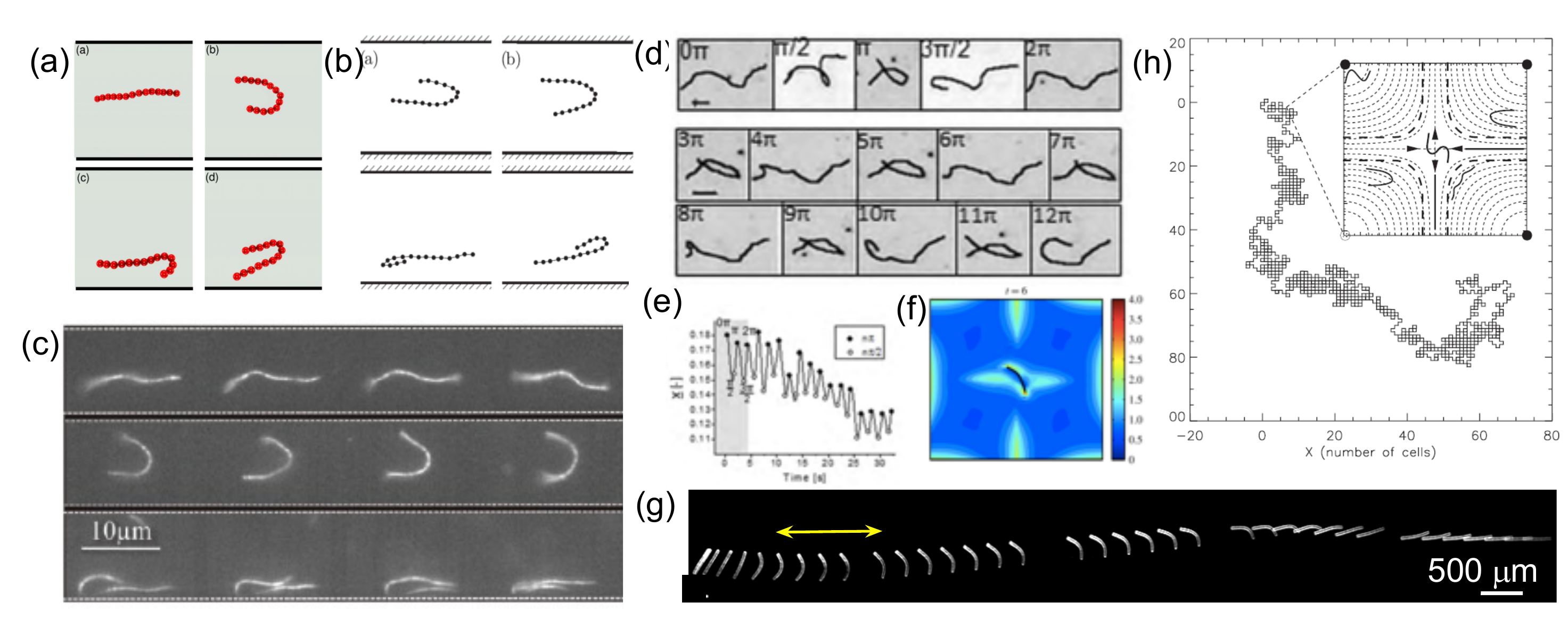}
		\caption{{\bf Fibers in complex flows.}
			{\bf (a), (b) \& (c) }Fiber
			shapes in Poiseuille flow. (a) \& (b) From bead-spring models, (a; \cite{Chelakkot2010}; b; \citep{Reddig2011}) and (c) from
			experiments using actin filaments, \citep{Steinhauser2012}.
			{\bf (d) \& (e)} Electrospun nanofilament in
			an oscillatory flow, \citep{Pawowska2017}. (d), Different
			morphologies and (e) normalized filament-wall distance during
			successive oscillations.
			{\bf (f)} Flexible fiber in cellular flow of a polymeric fluid,
			from simulations, \citep{Yang2017a}. The color scale indicates
			polymer stretching.
			{\bf (g)} Trajectory of a flexible polymeric fiber
			confined by the top and bottom walls of a microchannel, \citep{Cappello2018}.
			{\bf (h)} Simulations of the random walk of a
			flexible filament across a cellular flow, \citep{YS2007}. }
		\label{fig:ComplexFlows}
\end{figure}

\subsubsection{Poiseuille flow and interaction with bounding walls}
\label{subsubsecPoiseuille}

Channel (or Poiseuille) flows represent the majority of experimental
systems and, unlike linear shear flows, exhibit a shear gradient
across the channel width. This can affect fiber morphology when fiber
lengths are comparable to the channel dimensions. A first consequence
is that buckling instabilities (as described above) are not observed
and bending modes dominate. Typical fiber morphologies in Poiseuille
flows are reminiscent of "snaking" motions (\cite{Harasim2013})
(Fig.~\ref{fig:Shear}b) and can vary depending on the exact
positioning within the flow (\cite{Reddig2011, Chelakkot2010})
(Fig.~\ref{fig:ComplexFlows}a,b). Secondly, several groups have
demonstrated, using bead numerical models (\cite{Reddig2011,
	Chelakkot2010}) or experiments with actin filaments
(\cite{Steinhauser2012},Fig.~\ref{fig:ComplexFlows}c), that
semi-flexible polymers show strong cross-stream migration in shear
gradients and accumulate at preferred distances from channel
walls. Similar predictions have been made for flexible fibers
(\cite{Slowicka2013}).  Very recently \cite{Nakielski2015} observed
cross-stream migration using Brownian electrospun hydrogel
nanofilaments in an oscillatory flow (Fig.~\ref{fig:ComplexFlows}d)
and highlighted the dependance of migration dynamics on the initial
fiber positioning.

\subsubsection{Confined flows}
\label{subsubsecConfinedFlow}

Fiber transport in a Hele-Shaw cell (Fig.~ \ref{fig:flows}c)
represents a very different situation. Fibers with a diameter
comparable to the cell height are confined by the top and bottom walls
and now evolve in 2D in a plug flow.  Such confined microfabricated
fibers (sec. \ref{subsubsec:fiberfabrication}) have been observed to
bend into a C-shape facing backwards relative to the flow direction
(Fig.~\ref{fig:ComplexFlows}e). At first sight it is surprising that a
fiber transported at zero force in a plug-flow deforms, but this can
be explained as follows: friction with the top and bottom walls
decreases the fiber velocity compared to the average flow velocity and
the fiber can be seen as a moving obstacle. The fiber thus creates a
flow perturbation inducing an inhomogeneous force distribution along
the fiber. \cite{Cappello2018} have evaluated the force distribution
as a function of fiber confinement using modified Brinkman
equations. The resulting fiber shapes are found to be in good
agreement with experimental findings (\cite{Cappello2018}). The
coupling between fiber deformation, drift, and interaction with the
side walls (\cite{Nagel2018}) leads to complex trajectories and
finally a preferred alignment with the flow direction
(Fig.~\ref{fig:ComplexFlows}e).

\subsubsection{Cellular flows}
\label{subsubsec:cellular}

From their numerical study of fiber-flow interactions \cite{YS2007}
predicted that flexible fibers could move as random walkers across a
closed stream-line cellular flow (Fig.~\ref{fig:ComplexFlows}h).
Fiber buckling, occuring close to the stagation points, yields due to
its many degrees of freedom an effective randomness, allowing the
fiber to cross streamlines and to be transported across the flow. The
precise transport properties depend in a non-trivial way on fiber
flexibility and length, as has also been shown in experiments
(\cite{WQFLdR2010}). Numerical simulations by \cite{Manikantan2013}
addressed the role of Brownian fluctuations which were shown to
increase trapping of filaments in vortices and thus decrease transport
across the vortex array. \cite{Deng2015} have numerically confirmed
these findings. See \cite{BLY2011} and \cite{Young2009} for a
numerical study of two fibers interacting in such cellular flows and
\cite{Yang2017a} for transport in a polymeric cellular flow
(Fig.~\ref{fig:ComplexFlows}f.)

\section{SINGLE FIBERS MOVING UNDER EXTERNAL FORCES AND CONSTRAINTS}
\label{sec:SingleExtForces}

\subsection{Sedimenting Fibers}
\label{subsection:sedimentingfibers}

\begin{figure}
		\includegraphics[width=\textwidth]{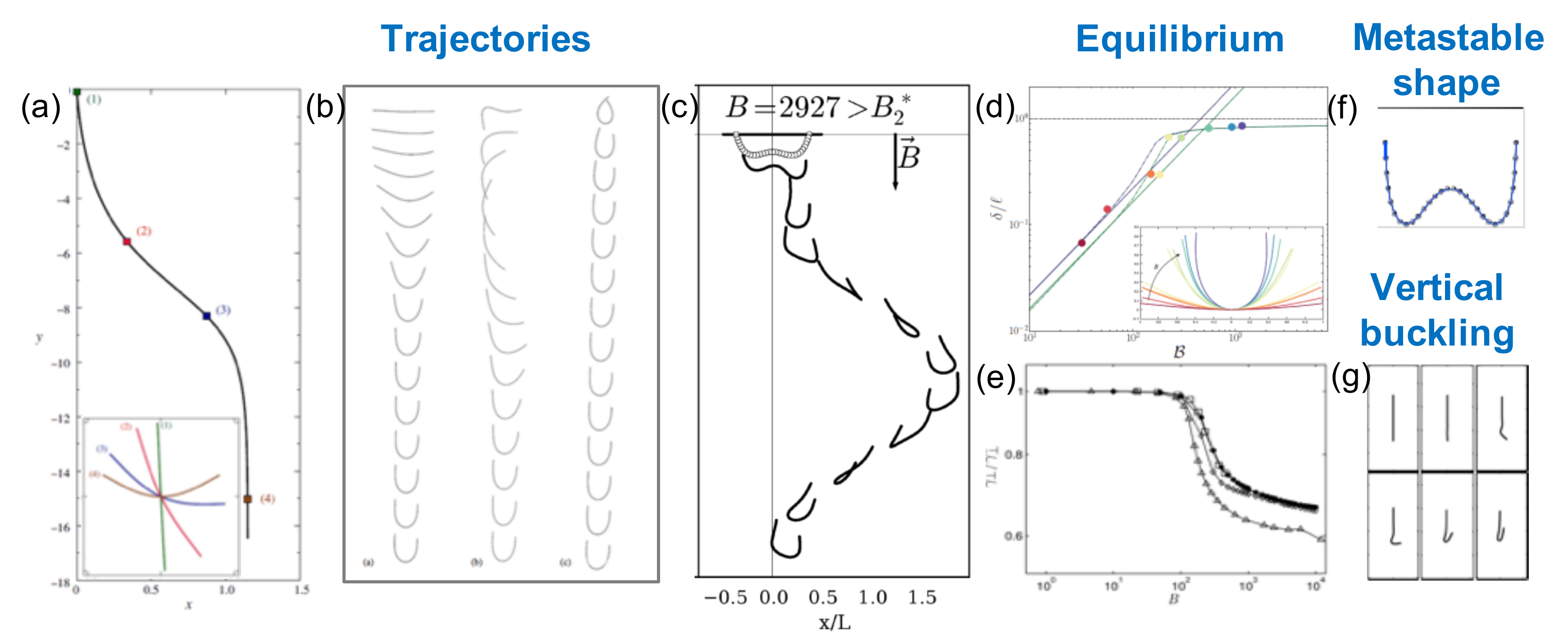}
		\caption{{\bf Sedimentation of flexible fibers.}
			{\bf (a)} Trajectory
			of the fiber center of mass, insert, fiber shapes during
			sedimentation, \citep{Li2013}.
			{\bf (b)} Chronophotographies of a
			fiber released with different initial conditions, \citep{Marchetti2018}.
			{\bf (c)} Complex trajectory of a fiber at
			very high $B$, \citep{Saggiorato2015}.
			{\bf (d)} Maximal
			deformation of the fiber at steady state as a function of $B$, \citep{Marchetti2018}.
			{\bf (e)} Evolution of viscous drag at
			steady state as a function of $B$, \citep{Delmotte2015}.
			{\bf (f)}
			Example of metastable configuration from bead-spring model, \cite{Cosentino2005}.
			{\bf (g)} Buckling instability of the
			leading fiber half from simulations using SBT, \citep{Li2013}. }
		\label{fig:BucklingSedAnchored}
\end{figure}

When flexible fibers are subject to external forces such as gravity,
the coupling between the elastic and viscous forces can induce complex
deformation and reorientation that strongly affects their macroscopic
transport. Sedimentation is the classical example (though
centrifugation would have similar consequences). The sedimentation of
flexible fibers was first studied from the theoretical point of view
by \cite{Xu1994}. Numerical studies have since been conducted using
bead models (\cite{Cosentino2005, Schlagberger2005, Delmotte2015,
	Saggiorato2015}) or slender-body theory (\cite{Li2013}). Only very
recently have experimental studies been performed
(\cite{Marchetti2018, Raspa2018}).

During sedimentation, fiber weight and viscous drag balance, and a
rigid fiber settles at constant velocity and orientation.
Hydrodynamic interactions along the fiber yield a viscous drag that is
smaller in the middle than at the edges. For a flexible fiber, this
inhomogenous drag deforms the fiber and leads to rotation.  The
theoretical work of \cite{Li2013} first explored how this affects
settling trajectories. As the drag depends on fiber orientation, the
rotation leads to a nontrivial coupling between horizontal and
vertical translations (Fig.~\ref{fig:BucklingSedAnchored}a). This
occurs until the fiber has reached an equilibrium configuration
characterized by a constant settling velocity and a bent shape, the
middle being perpendicular to gravity independently of the release
conditions
(Fig.~\ref{fig:BucklingSedAnchored}b). \cite{Saggiorato2015} have
explored numerically the behavior of very flexible and very heavy
filaments whose shape can become non-planar resulting in drift or
helical trajectories (fig.\ref{fig:BucklingSedAnchored}c).

The fiber shape at equilibrium depends on the relative magnitudes of the
gravitational force and the elastic restoring force. The relevant
control parameter is $B=F_g L^2/E$, with $F_g$ the gravitational force
(Fig.~\ref{fig:BucklingSedAnchored}d) (\cite{Cosentino2005,
	Delmotte2015, Saggiorato2015, Marchetti2018}).  For small $B$ the
steady velocity is close to the horizontal settling velocity of a
rigid fiber, while for large $B$, the fiber experiences large
deformations and the velocity is close to the vertical settling
velocity of a rigid fiber (Fig.~\ref{fig:BucklingSedAnchored}e)
(\cite{Delmotte2015}). \cite{Marchetti2018} recently confirmed this
experimentally and identified the intermediate regime as one of
continuous drag reduction through fiber reconfiguration.  Defining an
elasto-viscous number $\tilde\eta$ is not straightforward for this
problem as the viscous forces are function of fiber shape.


The case of a fiber settling while aligned with gravity is very
different: a inhomogenous fiber tension results from the inhomogenous
viscous drag, and is compressive in the leading-half and extensile in
the trailing part (\cite{Li2013}). Above a certain threshold of the
control parameter, buckling may occur in the leading half
(Fig.~\ref{fig:BucklingSedAnchored}f).


\subsection{Anchored fibers}
\label{subsection:anchoredfibers}

Passive anchored fibers, such as the primary cilium
(\cite{Young2012}), are found in biological systems, and can also form
spontaneously under flow conditions as in the formation of biofilm
streamers \citep{Drescher2013}.  In engineered micro-fluidic flow
geometries, fabricated anchored fibers can be used as flow sensors, or
conversely the micro-fluidic flows can be used to measure the bending
properties of unknown materials.  Depending on the anchoring and flow
orientation either buckling or bending can occur.

\subsubsection{Fiber buckling}
Using local SBT, \cite{GKSS2012} investigated the stability of elastic
fibers when held against an impinging linear or quadratic stagnation
point flow (Fig.~\ref{fig:Anchored}a). One fiber end is either clamped
or hinged at the wall (i.e. free to rotate with zero applied
torque). To identify critical values in $\tilde\eta$ for buckling
transitions they discretized the linearized dynamics equations and its
boundary conditions
and posed it as a generalized finite-dimensional eigenvalue problem
for growth rates. For the clamped filament a first unstable {\it
	bending} mode was identified for both flow fields. The next unstable
mode (in $\tilde\eta$) corresponds to buckling
(Fig.~\ref{fig:Anchored}a) with slightly different thresholds for the
two flows.  The hinged fiber is always unstable to rotation around its
base, and its higher modes correspond to buckling, having significantly
lower thresholds as compared to free fibers.


Experimental studies of fiber buckling in impinging flows are
difficult as even a slight misalignment between the fiber base and the
stagnation point leads to fiber bending (or rotation) instead of
buckling. However, situations where long fibers remain either
temporarily pinned, as observed for fibers flowing in rough fractures
(\cite{DAngelo2009}), or which experience a compressive flow only
locally, as observed during continuous fiber fabrication in a widening
microfluidic device (\cite{Nunes2013}), show buckling instabilities
similar to those of anchored fibers. In particular, \cite{Nunes2013}
show that such buckling can be used to form "crimped" microfibers with
a well defined morphology (Fig.~\ref{fig:Anchored}b).

\subsubsection{Fiber bending}
The deformation of anchored fibers by viscous flows is, in a number of
simple flow geometries, akin to a bending beam where viscous forces
play the role of a gravitational load. Microscopic beam bending
has been used by \cite{Duprat2015} to determine the bending
modulus of an elastomeric fiber fabricated using the UV projection
method introduced in Sect. \ref{subsubsec:fiberfabrication}. Here the
fiber was confined between the top and bottom channel walls and held
on both ends in a straight microchannel (Fig.~\ref{fig:fibers}l). In
this configuration flow takes place through the small gap above and
below the fiber and the viscous force applied can be determined using
a simple lubrication analysis.

\begin{figure}
		\includegraphics[width=\textwidth]{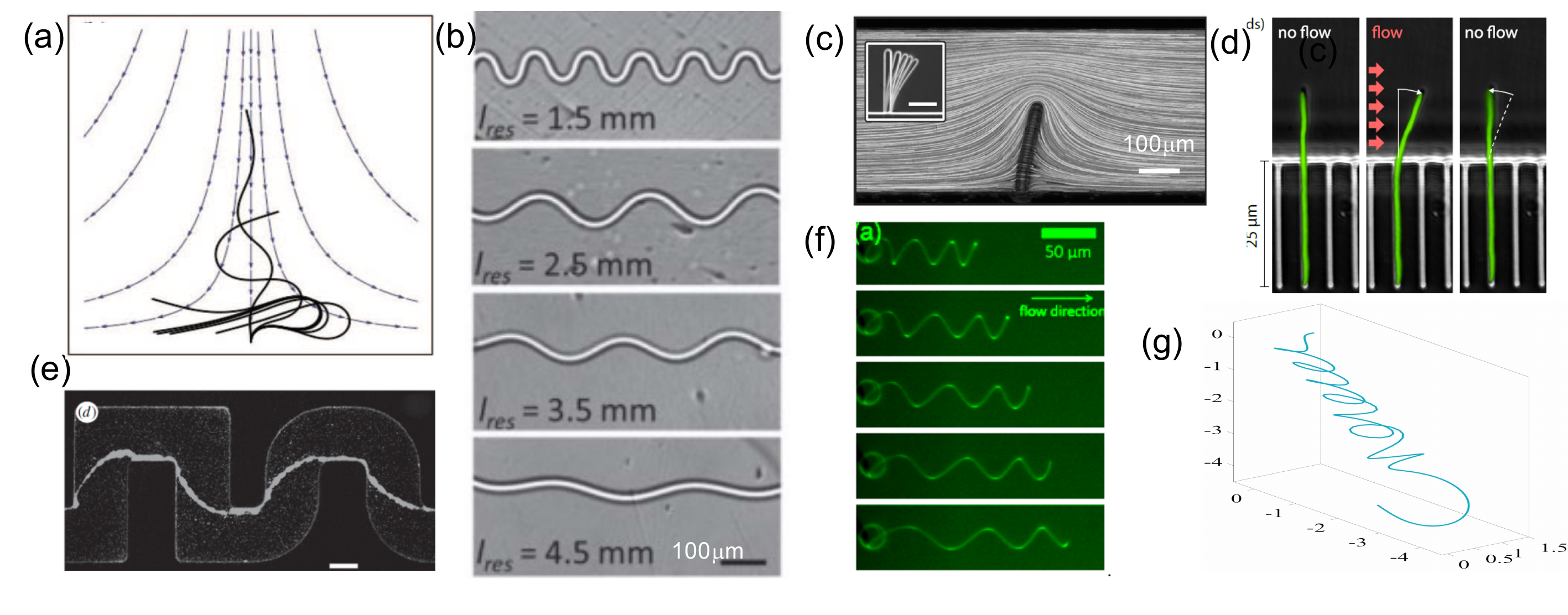}
		\caption{ {\bf Deformation of anchored fiber.}  {\bf (a)} Time evolution of a clamped filament buckling in an impinging flow,  \citep{GKSS2012}. {\bf (b)} Buckling  instability of a microfiber in a microchannel, \citep{Nunes2013}. {\bf (c) \& (d)} Deformation of objects  anchored perpendicular to the flow: a photopolymerized fiber  confined by top and bottom walls(c; \cite{Wexler2013}),  and single-cell {\it Escherichia coli}  grown in a lateral slit,(d; \cite{Amir2014}). {\bf (e)} Biofilm streamers grown in a curved  microchannel, \citep{Rusconi2010}.  {\bf (f)} Micro-helix deformed by a parallel flow, \citep{Pham2015}.  {\bf (g)} Clamped microtubule growing against a wall,  from E. Nazockdast. 
		}
		\label{fig:Anchored}
\end{figure}

The flow geometry is more complex when a confined fiber is attached to
one wall and so partially blocks a confined channel.  The resulting
flow profile has been visualized (Fig.~\ref{fig:Anchored}c) in
experiment by \cite{Wexler2013} using fibers attached to the wall of a
micro-channel and perpendicular to the flow direction. A theoretical
model (\cite{Wexler2013}) gave insight into the competition between
bending and leakage flow, showing good agreement with their
experimental results, and was also used to measure fiber rigidity. By
knowing its mechanical properties, such a fiber could be used as a
flow sensor in micro-fluidic devices.

\cite{Amir2014} studied the flow-induced bending of single-cell {\it
	Escherichia coli} growing from slits in the side-walls of a
micro-fluidic channel (Fig.~\ref{fig:Anchored}d).  By applying a
flow perpendicularly to the cells in their unconfined flow geometry,
the experimental set-up corresponds again to a simple bending beam
experiment where the force applied to the cell results from the
viscous friction of the fluid.  By estimating this viscous force and
using linear elasticity theory, the authors were able to estimate the
bending stiffness of the {\it E. coli} cells from the measured
deflection.  The simple hydrodynamic set-up gave values similar to
those obtained using much more costly techniques such as AFM
measurements.

Another biological example is the {\it primary cilium}. The primary
cilium is a non-motile and flexible hair-like protrusion from a cell into
the extracellular space and is found in a variety of vertebrate
cells. The primary cilium is believed to act as a
mechanoreceptor by bending in response to flow as is observed in
kidney tubule cells. Its dynamics has been investigated in a combined
experimental and theoretical study by \cite{Young2012}.

More complex flow geometries are experienced by biofilm streamers that
have been shown to grow into long, filamentous elastic structures
\citep{Rusconi2010, Rusconi2011} under flow. The shape of elastic
fibers within flows with curved streamlines has been determined
numerically by \cite{AGLRS2011} for fibers anchored, either hinged or
clamped, at given positions near a two-dimensional corner.  This work
shows that, due to tension and bending forces within the fibers, the
fibers do not align with the flow but rather cross flow
streamlines. This is in agreement with the experimental observations
from \cite{Rusconi2010,Rusconi2011} shown on
Fig.~\ref{fig:Anchored}e.

%

\subsubsection{Stretching of helical fibers}
The stretching by flow of very flexible anchored microhelices has been
studied by \cite{Pham2015}. Here the viscous drag represents a
distributed loading leading to a nonuniform pitch distribution
(Fig.~\ref{fig:Anchored}f). The overall helix extension is well
described by the FENE spring model, with the linear part simply
given by the ratio of viscous to bending forces. The micro-scale
helical ribbons with nano-scale thickness serve both as model systems
for bacteria flagella and as very sensitive flow sensors.

\subsection{Actuated fibers}

Biological cilia and flagella are of central importance to biophysical
processes like propulsion or embryogenesis (\cite{Bray2001,
	Hirokawa2006, Babu2013}). Both are elongated fibers anchored to
cellular membranes and which beat or rotate through the action of
internal molecular motors (like the cilia covering the surface of
micro-organisms; see \cite{Camalet1999}) or by rotating motors
anchored at their base (like bacterial flagella; see
\cite{Bray2001}). Their interactions with the surrounding fluid medium
lead to shape deformations to useful work like transport processes
like propulsion or mucus clearance in the lung. The study of anchored
filaments submitted to forces is also very important to understand the
coupling between cytoskeletal filaments and molecular motors that
occurs in the cell (e.g. \cite{TDDSJG2015,Shelley2016}) during
intracellular trafficking or to design micro-robots capable of motion
in a fluid.

\begin{figure}
		\includegraphics[width=\textwidth]{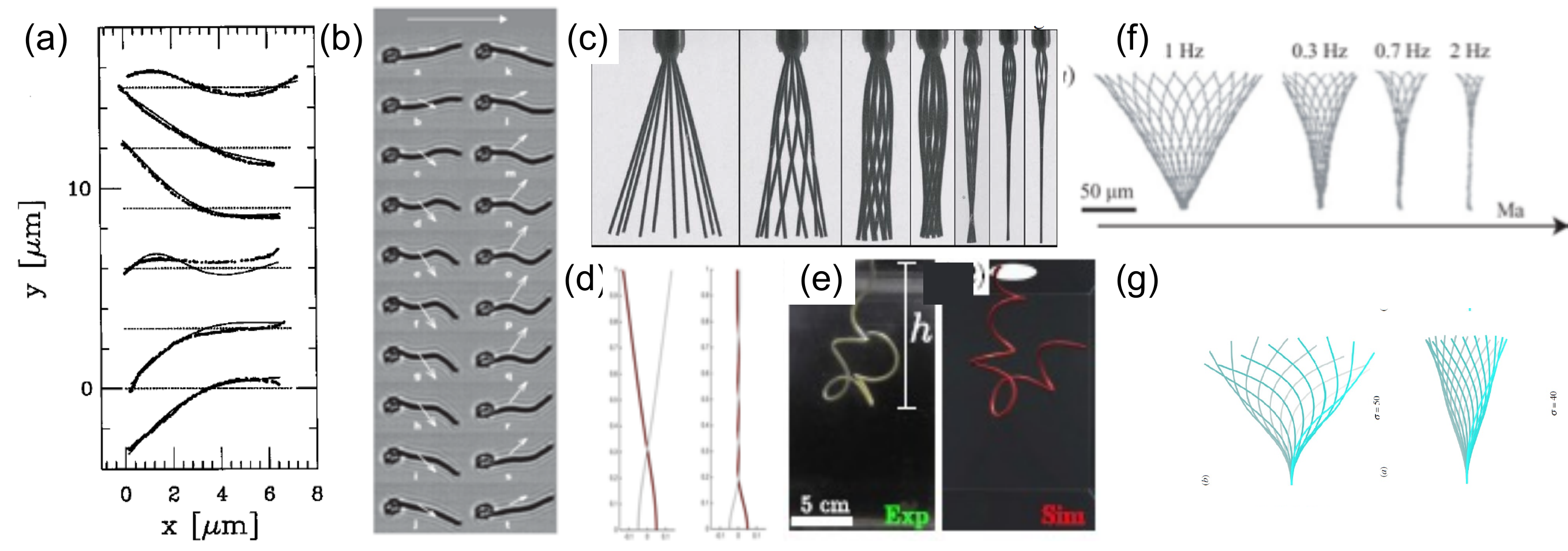}
		\caption{ {\bf Anchored and driven fibers.} {\bf (a) \& (b)} Fibers
			attached to a cargo. (a) A bead moved by an optical trap, \citep{Wiggins1998}.
			(b) A  magnetically actuated fiber propelling a red blood cell, \citep{Dreyfus2005}.  {\bf (c)} Rotation
			of a tilted elastic filament in a viscous bath at increasing
			rotation velocity, \citep{Coq2008}. {\bf (d)}
			Attractive interaction between two elastic fibers rotating around
			their own axes (left, experiments; right, numerical simulations,
			\citep{Man2017}. {\bf (e)} Buckling of a rotating flexible helix,
			\citep{Jawed2015}. {\bf (f) \& (g)} Shapes of
			beating fibers. (f) Magnetic fiber beating in a plane, \citep{Babataheri2011}; (g) A simulation of a fiber actuated at its
			free-end by a ``follower'' force aligned with the fiber,
			\citep{Canio2017}.
		}
		\label{fig:ActuatedExp}
\end{figure}

In the large literature on this topic, largely beyond the scope of
this review, fibers are usually anchored to a wall but in some studies
fibers are attached to an object small enough to be moved by the fiber
motions (i.e. flagella and bacteria). This was studied at the
micron-scale in two different experiments
(Figs.~\ref{fig:ActuatedExp}a-b).  Early to the field,
\cite{Wiggins1998} attached an actin filament to a bead that was
periodically driven by an optical trap. The bead oscillations induced
traveling waves on the fiber that generated propulsive force. In
another realization of this phenomenon \cite{Dreyfus2005} built an
artificial micro-swimmer from a chain of magnetic beads attached to a
red-blood cell - only here to break symmetry - and actuated the chain
with a time-dependent magnetic field.

In another class of studies, fiber motion is induced by the rotation
of the anchoring point. \cite{Wolgemuth2000} took inspiration from
biology to model the deformations induced by rotation of a fiber
around its axis in a viscous fluid. Rotation, twisting and bending
combine to yield different regimes of rotation, one of them having the
free end repelled from the axis. The interaction of two rotating fibers has been recently studied numerically and
experimentally by \cite{Man2017} (Fig.~\ref{fig:ActuatedExp}d), who
show that the induced rotational flows
tend to bend the fibers around each other. If a fiber is tilted, at
the anchoring point, relative to its rotation axis, the interaction
with the fluid leads to the collapse of the fiber onto the rotation
axis (Fig.~\ref{fig:ActuatedExp}c) while assuming a helicoidal shape
that generates a propulsive force (\cite{Manghi2006, Qian2008,
	Coq2008}). The rotation of a flexible helix can result in a
propulsive force sufficiently large to induce buckling
(Fig.~\ref{fig:ActuatedExp}e, \cite{Jawed2015}). The flexibility and
buckling of the flagellar hook at the base of the flagellum plays an
important role in the swimming of mono-flagellated bacteria where its
buckling yields changes in swimming direction
(\cite{Son2013}). Interactions between two weakly flexible helices
result in bundling as has been shown experimentally by \cite{Kim2003}.
As a model for cilia, \cite{Guo2018} studied the hydrodynamic coupling
of immersed elastic fibers driven at their base by an applied
torque. For two fibers their simulations and theory showed that
bistability of in-phase and out-of-phase oscillations can arise from a
coupling of elasticity and drive to flow.

Forced fibers can either be actuated locally or all along the
body. The latter case has been experimentally and numerically modeled
by magnetic colloid chains (\cite{Dreyfus2005, Gauger2006,
	Babataheri2011}).  The former has been addressed in different
experimental and numerical studies mainly with the forcing being
applied to the anchoring point (\cite{Wolgemuth2000, Wiggins1998,
	Manghi2006, Qian2008, Coq2008, Guo2018}), as already discussed, and
typically yield oscillations.  \cite{Canio2017} recently studied
tangential forcing at the fiber free-end as a model for the action of
motor proteins on microtubules, and demonstrated a bifurcation to
fiber beating via a buckling instability
(Fig.~\ref{fig:ActuatedExp}f-g).

\section{MICRO-MACRO COUPLING AND COLLECTIVE BEHAVIOR}
\label{sec:SuspDyn}

We have thus far focused on single fibers (or a few) interacting with
flows. However, in most industrial and natural settings flexible
fibers are mostly found collectively and in suspension. Understanding
the transport and rheology of fibrous suspensions at macroscopic
length scales is of significant importance.  In this chapter we review
studies on the relationship between microscopic dynamics of suspended
fibers and macroscopic rheology, as well as situations where fibers
act collectively in their interactions with flow as in ciliary beds or
in new flow control devices.

The main difficulty in studying fibrous assemblies or suspensions is
the complex and long-ranged nature of hydrodynamic interactions
(HIs). The importance of fiber interactions is defined by their
average separation distance.  In the dilute regime the fibers
separation distance is much larger than their length: $n L^3 \ll 1$
where $n$ is the fiber number density. In this limit HIs can be
neglected and for a suspension the ensemble behavior can be estimated
from the dynamics of a single fiber interacting with the
flow. Increasing the concentration causes HIs to be stronger. In the
semi-dilute regime of $nL^3>\approx 1$, the distance between fibers
becomes less than their length and both HIs and possibly steric
interactions become important. With further increase of concentration
the separation distance becomes comparable to the fiber radius and
rotational motions are hindered. For suspensions this can result in
formation of liquid crystalline phases (\cite{Doi1978}), which is
beyond the scope of this review.

\subsection{Rheology of fiber suspensions}

\subsubsection{Linear rheology for dilute non-Brownian suspensions}
The analysis of rheology for non-Brownian suspensions in oscillatory
background shear flows is very similar to the buckling instability
analysis in simple flows in \S \ref{subsubsec:buckling}. For small
perturbations of fiber shape from a straight rod, whose dynamics are
described by Eq.~\ref{StraightRod}, Eq.~(\ref{LSB_EB}) reduces to
\begin{align}
& \tilde{\eta} \left(w_t
-w \cos \tilde\omega t  \right)=
-w_{ssss}+2\bar{T}_s w_s+\bar{T} w_{ss},~~\&~~
\bar{T}=-\frac{\tilde{\eta}}{4}(s^2-\frac{1}{4})\cos \tilde\omega t,
\label{LinStabOsc}
\end{align}
where $w$ is the perturbation transverse to the fiber orientation in
the plane of shear, and $\tilde\omega$ is the oscillation frequency
normalized by the maximal shear rate $\dot\gamma$.

Perturbations will decay so long as $\tilde\eta<306.4$, the first
bifurcation to buckling for the constant shear case. The long-time
behavior will then be that of an oscillating rigid fiber which,
because of time-reversibility, will yield only a viscous response
under time average. For values of $\tilde\eta$ beyond the buckling
transition fibers under compressive stress will buckle, with that
buckling relieved in the reverse part of the cycle. Since elasticity
is relaxational throughout the oscillation it remains unclear whether
the long-time behavior supports persistent buckling. If not, the
rheology will remain purely viscous.


\subsubsection{Linear rheology of dilute Brownian suspensions}
The linear viscoelastic behavior of semi-flexible polymers has been
widely studied over the past two decades.  Examples of these systems
include worm-like polymers, cytoskeletal filaments and nanotubes.  A
detailed discussion of the literature in this limit is beyond the
scope of this review, and readers are referred to
\cite{Shankar2002}. Here we give a very brief summary relevant to
this review.

Rotational diffusion is the slowest time-scale for Brownian
semi-flexible polymers, and is given by $\tau_\text{rot}\sim \mu L^3/(
k_b T c)$. Another time-scale arises from curvature relaxation, given
by $\tau_\text{curv}=8\pi\mu L^4/Ec$.  
The ratio of bending to rotational
diffusion time-scales is $\tau_\text{curv}/\tau_\text{rot}=L/l_p \ll
1$.  The fastest relaxation time is associated with tensile forces. In
a rigid fiber the tension is set instantaneously. The presence of
thermal forces induces shape fluctuations in semi-flexible
polymers. Thus, the average end-to-end length of the polymer will be
less than its full contour length and will change under external
flows/forces. The tension time-scale is roughly defined as the time
$\tau_T$ over which the end-to-end distance of the polymer reaches
mechanical equilibrium under small external forces \cite{Shankar2002}.
The tension dynamics is more involved, but has been studied in depth
(\cite{Morse1998b, Everaers1999, Pasquali2001}). These studies find
that $\tau_\text{T}/\tau_\text{curv}=(L/l_P)^4 \ll 1$. Thus,
$\tau_\text{T} \ll \tau_\text{curv} \ll \tau_\text{rot}$.  This
separation of time-scales, and the linear response in small
deformation allows for analytical solutions of Eq.~(\ref{LSB_B})
in \S \ref{sec:BLSB}, which is outlined in \cite{Shankar2002}.

Having calculated the shape of the fiber, the time-dependent particle
stress tensor is found by the force moment integral along the
fiber: $\mathbf{S}=\int_0^L \mathbf{X} (s)\mathbf{f}(s)ds$, recalling
that $\mathbf{f}=\mathbf{f}^E+\mathbf{f}^B$ where
$\mathbf{f}^E=(T\mathbf{X}_s)_s-E\mathbf{X}_{ssss}$, and
$\mathbf{f}^B$ is given by Eq.~(\ref{FDT}). In dilute suspensions the
interactions between the fibers are ignored, and the ensemble average
stress reduces to $\mathbf{\Sigma}_F(t)=n \langle \mathbf{S} \rangle$.
\cite{Shankar2002} used this expression to compute
elastic and loss moduli in oscillatory shear flow for different values
of $L/l_p$. Fig.~\ref{fig:Susp}a gives a schematic presentation of the
tensile, bending, and thermal contributions to the time-dependent
shear modulus relaxation in a step strain deformation (see Fig.~2
in \cite{Shankar2002}). At very short times where
$t\ll\tau_\text{curv}$, the shear modules and its relaxation is
determined by tensile forces, while for $t \sim \tau_\text{curv}$ all
forces contribute equally to the rheology. Finally, at long times the
behavior is dominated by thermal forces; thus the long-time behavior
is viscous.

\subsubsection{Nonlinear rheology in shear flows}

Introducing flexibility to the fiber breaks time reversal symmetry of
Stokes equation and changes the behavior qualitatively. \cite{BS2001}
used local SBT to show that fiber buckling (see Fig.~\ref{fig:Shear}d)
gave rise to positive first normal stress differences, and that the
predicted threshold to buckling agreed well with the onset of positive
first normal stress differences in shearing experiments of nylon
fibers in glycerin \citep{Goto1986}.

\begin{figure}
	\includegraphics[width=\linewidth]{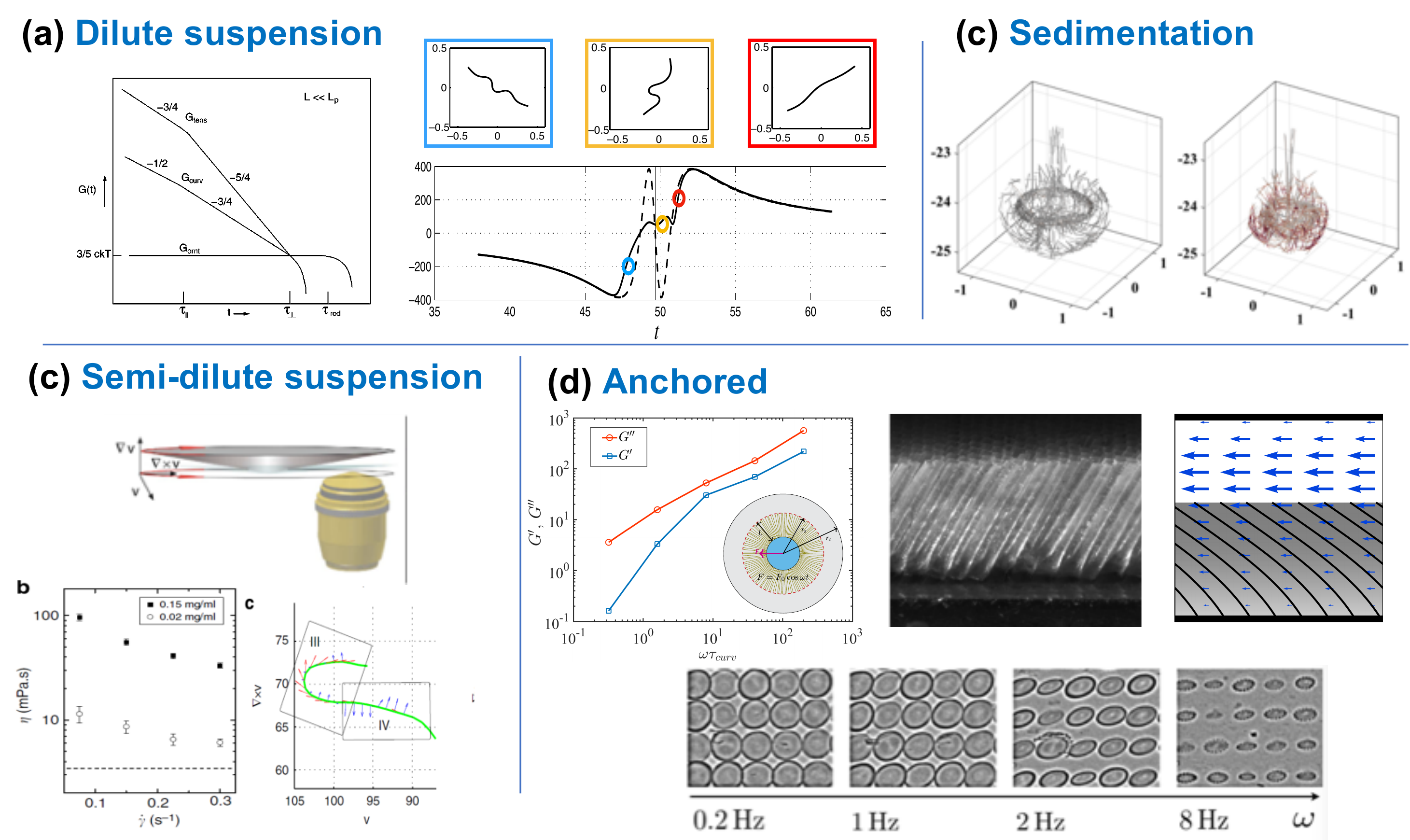}
	\caption{{\bf Collective behavior. }
		{\bf (a) Dilute suspensions:} Left, asymptotic contributions of
		tensile, bending, and thermal forces to shear modulus for
		semi-flexible polymers (\cite{Shankar2002}). Right, first normal
		stress difference during shear for, solid curve, straight filaments,
		and dashed curve, buckling filament (shapes on the top row)
		(\cite{TS2004}).
		{\bf (b) Semi-dilute suspensions.} Experimental
		set-up of \cite{Kirchenbuechler2014} combining a rheometer and
		confocal imaging to measure the shear viscosity of semi-dilute
		actin suspension (left) and image in 3D the filament shape
		(right).
		{\bf (c) Sedimentation:} Torus ring formed during the
		sedimentation of a cloud of $1024$ flexible fibers (more flexible in
		the right graph than in the left graph), (\cite{NRZS2017}).
		{\bf (d) Anchored fibers:} Top, left, rheology of a radially anchored
		flexible fiber shell, (\cite{NRZS2017}); Middle, array of tilted
		elastic fibers (\cite{Alvarado2017}); Right, a simulation of flow
		through a anchored bed of flexible fibers using a continuum model of
		an elastic fiber Brinkman medium (from D. Stein \& M. Shelley, in
		preparation). Bottom, trajectory of the tips of many rotating magnetic
		fibers with increasing frequency, (\cite{Coq2011}). }
	\label{fig:Susp}	
\end{figure}

\cite{TS2004} confirmed this observation in their nonlocal SBT
simulations.  Figure~\ref{fig:Susp}a shows the numerical predictions
of Tornberg \& Shelley for first normal stress difference induced by a
single flexible fiber in shear flow,
$N_1=\mathbf{S}_{xx}-\mathbf{S}_{yy}$, during one rotation of the
fiber (solid line). The results for a rigid rod are shown 
for comparison.  To understand the physical basis for this
behavior it helps to study the fiber's shape in three instances: (i)
the beginning of deviation from rigid rod, (ii) when the deviations
are maximized, and (iii) the return to a rigid rod behavior. These
instances are labeled with open circles and are shown with 
their corresponding shapes in Fig.~\ref{fig:Susp}a. We see
clearly that the deviation coincides with the buckling of fibers under
compressional stress ($\mathbf{p}\cdot
\mathbf{E}\cdot\mathbf{p}<0$).  As the fiber rotates to the
extensional quadrant ($\mathbf{p}\cdot \mathbf{E}\cdot\mathbf{p}>0$),
it is stretched back by the straining component of the shear flow to
its initial straight shape, and the corresponding stress approaches
that of a rigid rod.  Intuitively, buckling reduces the effective
hydrodynamic length (the end-to-end distance) of the fiber, decreasing
the stress magnitude in comparison with a rigid rod (recall that
$|\mathbf{S}|\sim L^3$).  Once the fiber moves to the extensional
flow quadrant, it straightens and its effective length
increase causes the stress magnitude to approach that of a rigid
rod.

\subsubsection{Shear rheology of semi-dilute suspension}
Several aspects of the rheology of flexible fibers in semi-dilute
regime remains poorly understood, due to the complexity of long-range
HIs. To our knowledge, there is no large-scale simulation of flexible
fibers in shear flow, with the exception of \cite{TS2004} who studied
the suspension of $N_F=25$ flexible fibers in oscillatory shear
flow. They observed that, at large enough shear rates, the fibers
deform during the period they are in compressional quadrants of the
flow. The fibers are then stretched back once they move to the
extensional quadrants. However, they don't entirely relax, perhaps
because of fiber-fiber interactions, and maintain strong buckling into
the ensuing periods. They also observe that the total stored elastic
energy is a strong function of the initial configuration of the
filaments, which the authors relate to the small number
of fibers in their simulations. For the same reason they did not
report on the rheology of the suspension. To our knowledge, there is
currently no prediction of shear viscosity and normal stresses {\it vs}
shear rate.

On the experimental front, \cite{Huber2014} studied the dynamics of
individual actin filaments in semi-dilute solutions of mixtures of
labelled and non-labelled filaments flowing in microchannels.  They
showed that steric hindrance reduces the tumbling rate significantly
in semi-dilute solutions as compared to isolated filaments. Entanglements between filaments can also lead
to anomalous or aborted tumbling events. The authors link the
microscopic filament dynamics to the shear thinning viscosity of the
semi-dilute solution via an effective diffusion coefficient.

The effect of entanglements on the three-dimensional deformations of
actin filaments in a semi-dilute solution was investigated
experimentally by \cite{Kirchenbuechler2014}. They used confocal
microscopy on a cone-and-plate counter-rotating shear cell and were
able to observe simultaneously microscopic filament dynamics and
macroscopic flow properties (Fig.~\ref{fig:Susp}b). They showed that
shear flow causes stretching and alignment of hair pin shaped
filaments explaining the observed shear thinning.

A very different behavior is observed by \cite{Perazzo2017} for
concentrated suspensions of very long fibers.  Using a microfluidic
fabrication method (Fig.~1i) they produced concentrated
suspensions of fibers with an aspect ratio greater than 100 and showed
that gelation, resulting from the formation of mechanical interlocking
of chains, is triggered by flow. This is also a
very simple way to produce biocompatible hydrogels.

\subsection{Sedimentation of fiber assemblies.}
The sedimentation of particles in fluids has been a major area of
research in suspension mechanics over the past several
decades \citep{Guazzelli2011}.  There have also been several
experimental \citep{Herzhaft1999, Metzger2005} and
numerical \citep{Butler2002, Saintillan2005, TG2006} studies on
sedimentation of suspensions of rigid fibers. The complex behavior in
sedimenting suspensions is due to HIs \citep{Ramaswamy2001,
	Butler2018}. For instance, sedimenting fiber suspensions form
inhomogeneous clusters resulting in enhanced sedimentation rates of
those clusters, in both numerical and experimental
studies (\cite{Metzger2005,Saintillan2005,TG2006}).
\cite{Manikantan2014} studied the effect of fiber flexibility
under small deformations on sedimentation using a continuum theory,
later followed by particle simulations \citep{Manikantan2016}. These
studies showed that flexibility can have both a stabilizing (no
clusters) and destabilizing (cluster formation) effect, depending on
the relative magnitudes of fiber flexibility and rotational diffusion.

Another interesting problem is the sedimentation of particle clouds.
Experiments and simulations have used rigid rods and spheres to show
that an initially spherical cloud evolves into a descending toroidal
shape reminiscent of a high Reynolds number vortex ring. The torus
eventually breaks into smaller clouds which evolve into tori
(\cite{Machu2001, Metzger2007,
	Metzger2005,Park2010}). \cite{NRZS2017}  used nonlocal SBT to
study clouds of $N_F=1024$ flexible fibers in a semi-dilute regime.
Their observations were in general agreement with the previous
experimental and computational studies on rigid rods
(Fig.~\ref{fig:Susp}c). They observed no qualitative effect of fiber
flexibility on the collective behavior of the cloud though flexibility
does result in changes in the shape of the torus ring (manuscript in
preparation). Fig.~\ref{fig:Susp}c shows the tori for two clouds with
fibers of differing effective rigidities (65 in ratio). In short,
increasing flexibility results in a smaller tori.

\subsection{Collective dynamics of anchored fibers.}
In many biological processes and industrial applications fiber
assemblies are attached to substrates and particles. Detailed
theoretical descriptions are still few, and a full accounting is
beyond the scope of our review, but we discuss several interesting
recent examples.

\cite{Alvarado2017} have recently shown how to construct soft flow
rectifiers by mounting dense beds of tilted flexible fibers onto the
walls of a flow channel (Fig.~\ref{fig:Susp}d, right panel).  When
fluid is driven in the direction of the fiber tilt, the fiber bed
depresses and facilitates flow. When fluid is pumped in the opposite
direction the bed rises and occludes flow. The mechanical responses
were quantified in a Taylor-Couette device, and rationalized
theoretically using a homogenous Brinkman approximation for fluid flow
through the bed.

In a related, but very different, work \cite{NRZS2017} studied
the rheology of centrosomal arrays of flexible microtubules in eukaryotic
cells. Using nonlocal SBT, they simulated the dynamics of flexible
fibers radially anchored to a sphere moving under an external oscillatory
force, $F=F_0\cos \omega t$ (schematic inset in
Fig.~\ref{fig:Susp}d, top left). At low frequencies
the response was purely viscous, and the fiber array was well-modeled
as a homogeneous Brinkman medium. At higher frequencies, the
force and velocity response moved out of phase, reflecting
viscoelasticity, and they used the phase lag to approximate
the elastic, $G^\prime$, and loss moduli, $G^{"}$
(Fig.~\ref{fig:Susp}d). For a single fiber $G^{"}/G^\prime$ has a
minimum at: $\omega_1=\tau_\text{curv}^{-1}$. A
minimum was also observed in the many-fiber simulation but at
$\omega \approx 25 \tau_\text{curv}^{-1}$. In other words, HIs between the
fibers decreased the relaxation time by a factor of $\approx
25$. An explanation is that the fibers create a poroelastic medium
whose fluid penetration length decreases with increasing fiber number.
Thus, bending modes with wavelengths above that of the
penetration length are prohibited, resulting in shorter effective
fiber length, and a much shorter relaxation time.

\cite{NRNS2017} used the same computational framework to study
\emph{pronuclear positioning}, an important transport process in
eukaryotic cells. They showed that ignoring HIs leads to
order-of-magnitude mispredictions of the necessary positioning forces.
They also showed that different proposed models for the active forces
gave rise to different cellular flows, and proposed flow measurement
as a tool to differentiate between active mechanisms.

Beds of actively beating cilia are common in biology, propelling
ciliated organisms, pumping fluid in the brain (\cite{FWBE2016}),
performing selective filtering (\cite{NawrothEtAl2017}), and clearing
mucus from the lungs. \cite{Mitran2007} carried out a 3D simulation of
propagating metachronal waves in rows of pulmonary cilia. They
considered a system where cilia sit primarily in a Newtonian viscous
fluid, while the tip moves within a viscoelastic fluid modeling a
mucus layer. They found that minimizing the work done by molecular
motors results in synchronized beating of cilia, and suggested that
HIs between cilia can lead to synchronized beating. \cite{DNM-NK2014}
use a regularized Stokeslet method to study the dynamics of an
infinite bed of driven cilia. They showed that metachronal waves not
only enhance fluid transport above the bed, but increases mixing of
fluid within it. Finally, in experiments \cite{Coq2011} investigated
the dynamics of extended arrays of artificial cilia driven by a
precessing magnetic field. Whereas the dynamics of an isolated cilium
was a rigid body rotation, collective driven beating results in a
symmetry breaking of the precession patterns. The trajectories of the
cilia are anisotropic and experience a significant structural
evolution as the actuation frequency increases (Fig.~\ref{fig:Susp}d,
bottom, left).

\section{SUMMARY AND OUTLOOK}
\label{sec:Summary}

As should be clear, flexible fibers interacting with
flowing liquids present a rich source of problems in fluid/structure
interaction. They also present complicated phenomena, which require
sophisticated experimental techniques to observe and measure, and
complicated theories through which to understand them.

There are many areas of open inquiry, among them the interactions of
fibers with complex media. Viscoelastic responses are typical of many
biological environments, such as the reproductive
tract \cite{Fauci2006} or inside of the cell \cite{Wirtz2009}.  There
is a developing literature on swimming of microorganisms in
viscoelastic fluids (\cite{EL2015,SA2015}).  Using a 2D slender-body
actuated elastica model, \cite{TG2017} recently studied the role of
body flexibility for undulatory swimming in viscoelastic fluids.
There have been few if any studies of flexible fibers interacting with
complex flows of complex fluids; Recent simulations
of \cite{Yang2017a} (Fig.~\ref{fig:ComplexFlows}f) of fiber transport
by cellular viscoelastic flow is the first of which we are aware.  The
fundamental theoretical difficulty is the necessity of evolving bulk
elastic stresses via transport nonlinearities. This makes
fluid-structure problems, much less those with multiple elastic
bodies, very challenging for viscoelastic flows and morally equivalent
to those for the Navier-Stokes equations. On the experimental side, it
remains challenging to synthesize complex fluids with
well-characterized (and simple!) rheological responses.

New kinds of mathematical coarse-grained descriptions need to be
developed to describe collective behavior of flexible fibers in
fluids, especially when hydrodynamic interactions are strong. One
regime where progress is being made is when the fibers can be
considered as well-aligned.  In recent work, Stein \& Shelley (in
preparation) have developed a continuum Brinkman-type model that
captures the anisotropic drag from elongated flexible structures to
compute the flow feedback to the bending and tensile response of a
porous elastic medium
(\cite{MoeendarbaryEtAl2013,SCLG2015}). Figure~\ref{fig:Susp}d (top
right) shows the model's result in simulating a soft flow rectifiern
by the bending of a bed of tilted and anchored elastic fibers
(Fig.~\ref{fig:Susp}d (top middle) from \cite{Alvarado2017}).

\section*{ACKNOWLEDGMENTS}
E.N. and M.J.S. acknowledge the support of the National Science Foundation, the National Institutes
of Health, and the Department of Energy. A.L. acknowledges support from European
Research Council Consolidator Grant 682367 under the European Union's Horizon 2020
program.

%
\section*{LITERATURE\ CITED}

\end{document}